\documentclass{ws-p8-50x6-00}
\def \half{\frac{1}{2}}
\def \eps{\epsilon}
\def \veps{\varepsilon}
\def \d{\partial}
\def \lam{\lambda}
\def \non{\nonumber}
\def \cLa{{\cal L}_1^{\pm 1}}
\def \cLb{{\cal L}_2^{\pm 1}}
\def \cLc{{\cal L}_3^{\pm 1}}

\renewcommand{\(}{\left(}
\renewcommand{\)}{\right)}
\newcommand{\df}[2]{ \frac{\partial {#1}}{\partial {#2}} }

\newcommand{\I}{\mbox{\rm i}}

\begin{document}

\title{Instabilities of Relativistic Stars}

\author{K.~D.~Kokkotas and J.~Ruoff}

\address{Department of Physics, Aristotle University of Thessaloniki,
Thessaloniki 54124, Greece\\ E-mail: kokkotas@astro.auth.gr;
ruoff@astro.auth.gr}


\maketitle

\abstracts{ Recent developments on the rotational instabilities of
  relativistic stars are reviewed. The article provides an account of
  the theory of stellar instabilities with emphasis on the rotational
  ones. Special attention is being paid to the study of these
  instabilities in the general relativistic regime. Issues such as the
  existence relativistic r-modes, the existence of a continuous
  spectrum and the CFS instability of the w-modes are discussed in the
  second half of the article.}

\section{Introduction}

The oscillations and instabilities of relativistic stars gained a
lot of interest in the last decades because of the possible
detection of their associated gravitational waves. It is not
impossible that every compact star in a specific period of its
life will undergo an oscillatory phase, in which it becomes
unstable. Only the dynamical instabilities were thought to be a
relevant source of gravitational waves, whereas instabilities due
to dissipation mechanisms were believed to be only of academic
interest. This belief has been dramatically altered during the
last five years after it was discovered that for a specific class
of rotational perturbations, the so called
r-modes\cite{Nils98,FM98}, the instability due to gravitational
radiation\cite{Chandra70a,Chandra70b,FS78a,FS78b} has the
potential of being a prime source for gravitational waves. It was
subsequently shown that this instability has many interesting
astrophysical implications, which attracted the attention of both
relativists and astrophysicists.

In the first section of this review we shortly describe the
various stellar instabilities and how they operate, for a more
detailed review one can refer to a recent article by
Stergioulas\cite{Nick98}. The second section is devoted to the
r-mode instability with the focus on its application to
gravitational-wave research and astrophysics, for more details see
a recent review by Andersson and Kokkotas\cite{AK01}. The last
section is devoted to an analysis of new features brought about by
the relativistic treatment of the problem. Questions on the
existence relativistic r-modes and a continuous spectrum are
discussed in detail. A new result that is presented is the
existence of the w-mode instability in ultra-compact stars,
i.e.~the instability due to the existence of an ergosphere of the
spacetime or w-modes\cite{KRA02}.

\section{Stability of Relativistic Stars}

\subsection{Non-rotating stars}
There are many different types of instabilities which can operate
in a relativistic star. The most familiar instability is due to
the existence of a maximum mass beyond which the star collapses.
In other words it becomes unstable with respect to radial
perturbations. The points of instability can be easily found by
constructing a sequence of stellar models for a particular
equation of state (EOS), parametrized for example by the central
density $\eps_c$. One can thus obtain the Mass-Radius diagram,
where the extrema signal the transition from stability to
instability and vice-versa. Within a perturbative approach, the
stability analysis with respect to radial pulsations is reduced to
a second-order Sturm-Liouville type equation\cite{Chandra64,KR01}
with eigenvalue $\sigma^2$. For $\sigma^2>0$ the eigenfrequency
$\sigma$ is real and the mode is stable while for $\sigma^2<0$ the
eigenfrequency $\sigma$ becomes imaginary and the mode is
unstable.  The case $\sigma^2=0$ marks the onset of the
instability. For the special case of uniform density stars,
general relativity predicts that the star becomes unstable with
respect to collapse if $M/R>8/9$.


For {\em non-radial perturbations} of non-rotating stars a
standard criterion for mode stability is the sign of the so called
Schwarschild discriminant
\begin{equation}
  {\cal A}_s={d p\over d r}-\left({\partial p \over \partial
      \eps}\right)_s {d \eps \over d r},
\end{equation}
where $r$ is the radial coordinate, $p$ the pressure of the fluid
and $\eps$ the total energy density. The condition ${\cal A}_s>0$
everywhere inside the star is necessary for stability. If ${\cal
  A}_s<0$ somewhere inside the star, the $g$-modes become unstable
with respect to convection\cite{Unno}. The stability of
non-radially oscillating stars can also be studied via
perturbation theory. One can derive a system of perturbation
equations for the fluid and the spacetime\cite{Thorne67}. For
relativistic stars the frequency of oscillation modes is no longer
real valued because the non-radial oscillations generate
gravitational radiation which dissipates energy from the star.
Instead the frequency is complex ($\sigma=\sigma_R+i sigma_I$) and
the imaginary part of the frequency describes the damping or the
growing time of a mode. Unstable modes are characterized by $\tau
<0$. For non-rotating relativistic stars with ${\cal A}_s>0$ no
unstable modes have been found, although a rigorous proof of
stability is still lacking.

\subsection{Rotating Stars}

The size of a rotating configuration is limited by the condition
that the centrifugal acceleration should not exceed the
centripetal acceleration of gravity. This leads to the following
upper limit for the angular velocity $\Omega$ in terms of the mean
density ${\bar  \eps}$, established already a century
ago\cite{Poincare}:
\begin{equation}
  \Omega^2 \leq 2\pi G{\bar \eps}.
\end{equation}
The limiting rotation frequency, $\Omega_K$, is called the {\em
Kepler limit}. Although derived in Newtonian theory using
simplifying assumptions, this relation is approximately correct
for relativistic systems as well\cite{HZ89,HSB95}.

For uniformly rotating stars, stability against {\em axisymmetric
perturbations} can be determined, via turning points along a
constant angular momentum sequence \cite{FIS88}. An initially
uniformly rotating relativistic star under the influence of such
an axisymmetric instability becomes differentially rotating. Since
the viscosity is the main mechanism for the redistribution of
angular momentum within the fluid, the timescale for the
development of the instability is set by viscous dissipation.

 {\em Rotational instabilities} of stars arise from non-axisymmetric
perturbations of the form $e^{im\phi}$, where $\phi$ is the
azimuthal angle. The bar mode ($m=2$) is known to be the fastest
growing unstable mode. There are two types of rotational
instabilities: the {\em dynamical} and the {\em secular} ones.
\begin{itemize}

\item The {\bf dynamical instability} is driven by hydrodynamical and
  gravitational forces and develops on an extremely short timescale,
  the dynamical timescale.  For neutron stars it is of the order of
  milliseconds.  Of special interest is the so called {\em bar mode
    instability} ($l=m=2$) which deforms the star into in a bar shaped
  structure within approximately one rotational period. The bar mode
  instability is an extremely efficient mechanism of producing
  gravitational waves\cite{HCS94,SHC96,Liu01}.

\item The {\bf secular instabilities} are driven by dissipative
  processes such as viscosity or gravitational radiation. They develop
  on much longer timescales, which are of the order of the dissipation
  time (for example the time required for viscosity to redistribute
  the angular momentum), i.e.~tenths of seconds or even more.

  The {\em viscosity driven instability} plays hardly any role in
  neutron stars because it occurs only in stellar models with much
  stiffer EOS than those considered to be realistic\cite{BFG96,SL96}.

  Chandrasekhar\cite{Chandra70a,Chandra70b} first proved that
  gravitational radiation can drive Maclaurin spheroids, that are
  rigidly rotating uniform density spheroids, unstable. This
  instability causes the Maclaurin spheroids to evolve into a
  stationary but non-axisymmetric configuration, the Dedekind
  ellipsoid\cite{DL77}.  This interesting result was put on a rigorous
  footing by Friedman and Schutz\cite{FS78a,FS78b,Friedman78a}, who
  also proved that this instability is generic and that {\em all
    rotating perfect fluid stars are unstable with respect to
    gravitational radiation}.
\end{itemize}

\subsection{The CFS instability}

The mechanism for the gravitational-wave instability can be understood
in the following way\cite{FS78a,FS78b}. Consider first a non-rotating
star. Then the mode-problem leads to eigenvalues for $\sigma^2$,
i.e.~one obtains equal values $\pm |\sigma|$ for the forward and
backward propagating modes (corresponding to $m=\pm|m|$). These two
mode branches are affected by rotation in different ways. The
retrograde mode will be dragged forward by the stellar rotation. If
the star spins sufficiently fast, this mode will appear moving forward
in the inertial frame, but still backward in the rotating frame. Thus,
an observer at infinity sees gravitational waves with positive angular
momentum emitted by this retrograde mode, but since the perturbed
fluid rotates slower than it would in absence of the perturbation, the
angular momentum of the mode itself is negative. The emission of
gravitational waves consequently makes the angular momentum of the
mode increasingly negative thus leading to an instability.

From the above, one can easily conclude that a mode will be
unstable if it is retrograde in the rotating frame (an observer
rotating with the star and observing a frequency $\sigma$ for the
mode) and prograde for a distant observer measuring a mode
frequency $\sigma-m\Omega$, i.e. if
\begin{equation}
  \sigma(\sigma-m\Omega) < 0. \label{criter1}
\end{equation}
For the high frequency ($f$ and $p$) modes this is possible only
for extremely high values of $\Omega$ or for quite large $m$. In
general, for every mode there will always be a specific value of
$m$ for which the mode will become unstable.

This class of {\em frame-dragging instabilities} is usually
referred to as Chandrasekhar-Friedman-Schutz (CFS) instabilities.
It is easy to see that the CFS mechanism is not unique to
gravitational radiation. Any radiative mechanism will have the
same effect.

A new result described in Section 4.3 is that the CFS mechanism is
not only active for fluid modes but also for the {\em spacetime}
or the so called {\em w}-modes\cite{KRA02}.

\begin{figure}[h]
\hbox to \hsize{\hfill \epsfysize=5cm \epsffile{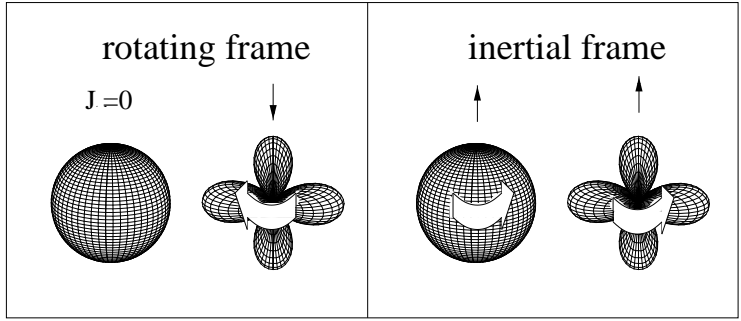}
\hfill} \caption{A schematic illustration of the conditions under
which the CFS instability is operating. A perturbed star can be
viewed as a superposition of a uniformly rotating background and a
non-axisymmetric perturbation. A mode is unstable if it is
retrograde according to an observer rotating with the fluid
(left), but appears prograde for an observer at infinity (right).}
\end{figure}

\subsection{Dissipative Effects}

Dissipation is important for the stability/instability of rotating
stars. In general dissipation due to the fluid viscosity tends to
suppress the instability and thus stabilizes stars that do not
rotate fast enough. The typical fluid viscosities that have been
used for the study of unstable $f$ and $r$-modes are the {\em
shear} and {\em
  bulk} viscosity\cite{IL91}. The effects of the dissipation on the
evolution of a perturbation are most easily studied by considering the
energy of the perturbation as measured in the rotating frame
\begin{equation}
  E={1\over 2} \int \left[\rho \delta u^\alpha \delta u^*_\alpha +
    {\delta p \ \delta \eps^* \over \eps} + \delta \Phi \delta \eps^*
  \right] dV, \label{energy}
\end{equation}
where $\delta u$ and $\delta \Phi$ are the perturbations of the
fluid velocity and gravitational potential. In the absence of
dissipation the energy of the mode is conserved. For a mode which
is subject to dissipation, the eigenfunction is $\sim
\exp[i(\sigma + i/\tau)t]$, where $\sigma$ is the real part of the
mode frequency and $1/\tau$ its imaginary part. The problem of
estimating the stability of a mode is reduced to determining the
sign of the imaginary part of the frequency. The imaginary part of
the frequency is the ratio between the energy of the mode and the
rate at which a specific dissipation mechanism dissipates energy
from the mode, i.e.
\begin{equation}
  {1\over \tau} = - {1 \over 2 E} {d E \over dt}.
  \label{damp_time}
\end{equation}
One now has to compare the damping/growth times due to the various
mechanisms for energy dissipation in a given neutron star model
\begin{equation}
  {1\over \tau} = {1 \over \tau_{\rm gw}} + {1 \over \tau_{\rm
      shear}} + {1\over \tau_{\rm bulk}}. \label{instabil1}
\end{equation}
Rough estimates of the the various viscous mechanisms are given
below.
\begin{itemize}
\item {\em Shear Viscosity} is the main dissipation mechanism for
low temperatures ($T \leq 10^9K$) and it arises from the momentum
transport due to neutron-neutron scattering. The dissipative
energy loss in the mode energy is given by
  \begin{equation}
    {dE \over dt} = 2\int \eta \ \delta \sigma^{\alpha \beta} \ \delta
    \sigma^*_{\alpha \beta} \ d V, \label{shear1}
  \end{equation}
  where $\sigma_{\alpha \beta}$ is the shear tensor and the
  thermodynamic function $\eta$ represents the shear viscosity of
  the fluid. A typical value for neutron stars is\cite{Sawer89,CL92}
  \begin{equation}
    \eta=2 \times 10^{18} \eps_{15}^{9/4}T_9^{-2} \mbox{ g  cm$^{-1}$
      s$^{-1}$} .
  \end{equation}

\item {\em Bulk Viscosity} dominates at high temperatures ($T \ge
10^9K$) when matter becomes transparent to neutrinos. As the
oscillating fluid undergoes compression and expansion, the weak
interaction requires a relatively long time to re-establish
equilibrium. This creates a phase lag between density and pressure
perturbations which results to a large bulk
viscosity\cite{Sawer89}. Typically the mode energy lost through
bulk viscosity is radiated away by neutrinos.  This energy loss
can be estimated by:
  \begin{equation}
    {dE \over dt} = \int \ \zeta \ \delta \sigma \ \delta \sigma^* \
    dV, \label{bulk1}
  \end{equation}
  where $\delta \sigma$ is the expansion associated with the mode and
  $\zeta$ the coefficient of the bulk viscosity, which for mode
  frequencies (in the rotating frame) has the value
  \begin{equation}
    \zeta = 6\times 10^{25} (\sigma+m\Omega)_{\rm 1Hz}^{-2} \
    \eps^2_{15} \ T^6_9 \ \mbox{ g cm$^{-1}$ s$^{-1}$} .
  \end{equation}

\item {\em Gravitational Radiation} dissipates energy from an
  oscillating star at a rate estimated via the post-Newtonian
  multipole-formula\cite{Thorne80} as
  \begin{equation}
    {dE \over dt} = -\sigma (\sigma + m\Omega) \sum^\infty_{l=2} N_l
    \sigma^{2l} \left(|\delta D_{lm}|^2+|\delta J_{lm}|^2\right),
    \label{gw1}
  \end{equation}
  where
  \begin{equation}
    N_l= {4\pi G \over c^{2l+1}} {(l+1)(l+2) \over l(l-1)[(2l+1)!!]^2}.
  \end{equation}
  The radiation emitted is the sum of the mass $\delta D_{lm}$
  and current $\delta J_{lm}$ multipole contributions given by
  \begin{eqnarray}
    \delta D_{lm} &=&\int \delta \eps \ r^l \ Y^*_{lm} d V, \\
    \delta J_{lm} &=&{2\over c}\left({l\over l+1}\right)^{1/2}\int r^l
    (\eps \delta \vec{u} + \vec{u}\delta\eps) \cdot
    \vec{Y}_{lm}^{B*} d V,
  \end{eqnarray}
  where the $Y_{lm}$ are the spherical harmonics and
  $\vec{Y}_{lm}^{B*}$ the magnetic type
  multipoles\cite{Thorne80,LOM98}.
\end{itemize}

By estimating the energy of the mode using equation (\ref{energy}) and
the energy dissipation via (\ref{shear1}), (\ref{bulk1}) and
(\ref{gw1}) one gets all the components needed in equation
(\ref{instabil1}). For a given EOS and a given stellar model the sign,
which signals if the mode is growing or not, of equation
(\ref{instabil1}) is a function of the {\em rotational period} and the
{\em temperature} of the star. Figure 2 shows the instability window,
obtained in this way, of the $l = m = 2$ $r$-mode.

\section{$r$-modes}

The oscillation patterns of a star are classified according to the
dominant restoring force that tries to push a displaced fluid
element back into its equilibrium position. The two main families
of modes, which exist for any type of stars (rotating or not), are
those whose restoring forces are pressure ($p$-modes) and buoyancy
($g$-modes). Rotation not only shifts the spectra of these modes
but additionally gives rise to a new type of restoring force, the
Coriolis force, with an associated new family of {\em rotational }
modes. In general these are called {\em inertial} modes and are
known in geophysics as Rossby waves. Of special interest to us is
the quadrupole $r$-mode with $l=2$, $m=2$.

Inertial modes are primarily velocity perturbations. In the slow
rotation limit, the velocity field of the $r$-mode is given by
\begin{equation}
  \delta \vec{u} \sim \alpha R \Omega \left({r \over R}\right)^l
  \vec{Y}^B_{ll}e^{i\sigma t},
\end{equation}
where $\alpha$ is the mode amplitude and $R$ the radius of the star.
In a first approximation the fluid elements have no radial
displacement and move on paths which are approximately ellipses with
$\theta$-dependent eccentricities
\begin{equation}
  {\xi_\theta^2 \over \sin^2\theta} +{\xi_\phi^2 \over \sin^2\theta
    \cos^2 \theta}\sim \alpha^2 R^2,
\end{equation}
see for example Fig.~1 in\cite{AK01}. The associated density
perturbations are small for low rotation rates $\delta \rho \sim
{\cal O}(\Omega^2)$ and at the Newtonian level, the mode frequency
in the rotating frame of reference is
\begin{equation}
  \sigma = {2m\Omega \over l(l+1)}.
\end{equation}
Using the criterion for CFS stability given by equation
(\ref{criter1}) we get
\begin{equation}
  \sigma(\sigma-m\Omega) =-{2(l-1)(l+2)m^2\Omega^2 \over l^2(l+1)^2}
  < 0,
\end{equation}
which implies that the $r$-mode is unstable for any rotation rate of
the star. This was first pointed out 4 years ago\cite{Nils98,FM98} and
immediately created considerable excitement because of the
astrophysical implications this instability would have for the spin
evolution of newly born neutron stars and for the search for
gravitational waves.

The main gravitational wave output from $r$-modes is not due to
disturbances in the star's fluid density (mass multipoles, $\delta
D_{lm}\sim \Omega^2$ ) but originates from the time dependent mass
currents (current multipoles, $\delta J_{lm}\sim \Omega$). {\em
This is the gravitational analogue of magnetic multipole
radiation} and is a rare case among the expected astrophysical
sources of gravitational radiation.

\subsection{Instability window}

One can estimate the growth time of the instability using equation
(\ref{damp_time}) together with the viscous damping times. In a
first approximation one obtains
\begin{eqnarray}
  \tau_{\rm gw} &\sim& 47 \ M^{-1}_{1.4} \ R^{-2l}_{10} \
  P^{2l+2}_{-3} \ {\rm sec} , \\
  \tau_{\rm sv} &\sim& 7\times 10^7 \ M_{1.4}^{-5/4} \ R_{10}^{23/4}
  \ T_9^2 \ {\rm sec}  , \\
  \tau_{\rm bv} &\sim& 3\times 10^{11} \ M_{1.4} \ R_{10}^{-1} \
  P_{-3}^2 \  T_9^{-6} \ {\rm sec} .
\end{eqnarray}
From the above values one can easily show that for high temperatures
($T \geq 10^{10}$K) the bulk viscosity suppresses any oscillations,
while the shear viscosity dominates for temperatures below 10$^7$K.
Still, as shown in Fig.~2, for temperatures between 10$^7$ and
10$^{10}$, there is a broad window in which the instability can
operate\cite{LOM98,AKS99}.

In particular these results imply that for temperatures of the
order of 10$^9\,$K a small rotation rate of the order of 5\% of the
Kepler frequency $\Omega_K$ is already sufficient to drive the
star unstable.

\begin{figure}[t]
  \hbox to \hsize{\hfill \epsfysize=6cm \epsffile{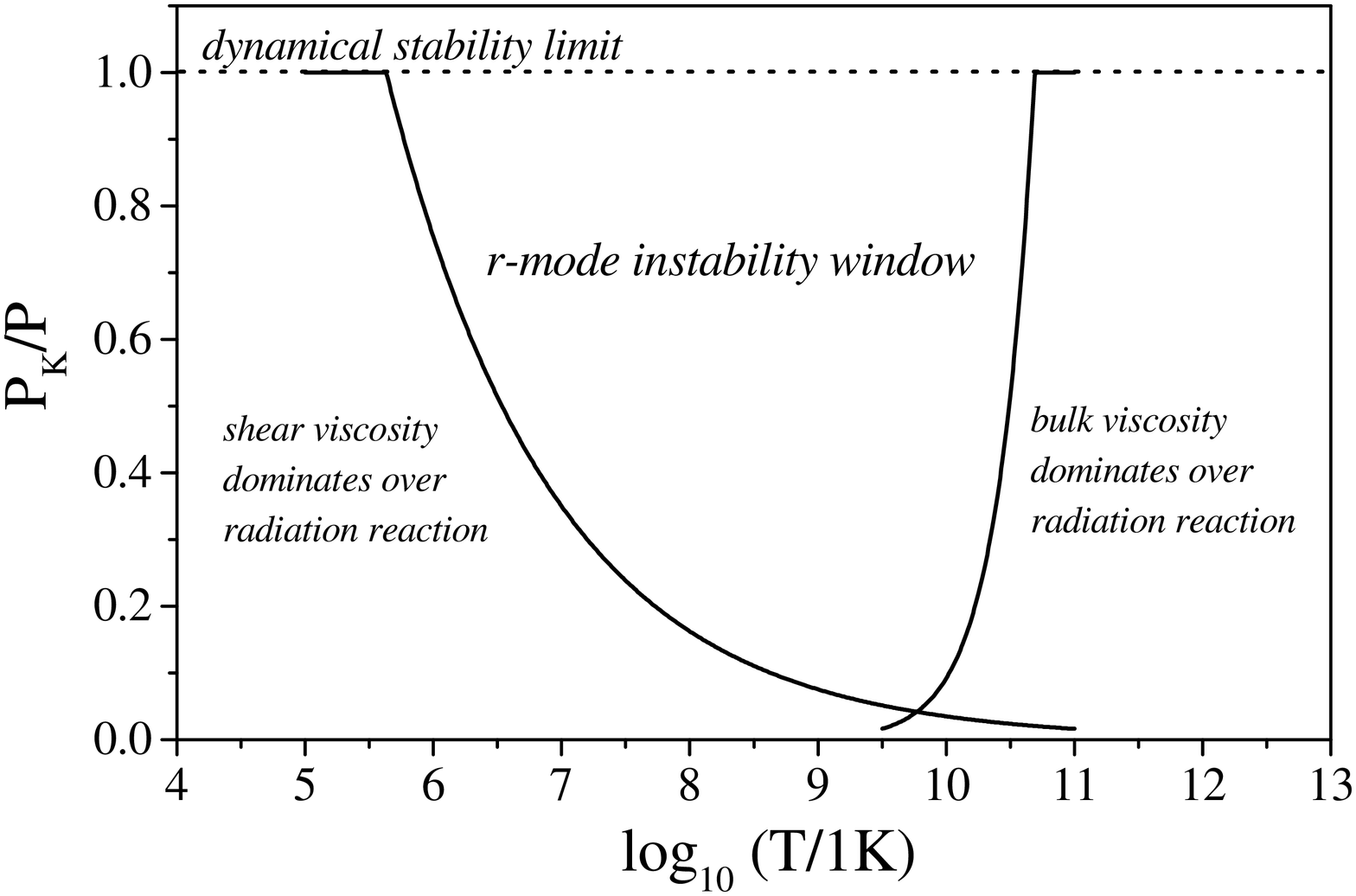} \hfill}
  \caption{The instability window for the $l=m=2$
    $r$-mode of a typical neutron star ($R=10$ km and $M=1.4 M_\odot$
    and Kepler period $P_K \approx 0.8$  ms). }
  \label{win1}
\end{figure}

The growth time due to the mass multipole radiation of the r-modes is
about four orders of magnitude longer and the corresponding
instability window is considerably smaller\cite{AK01}. The picture is
similar for $r$-modes with higher $l$.

In order to obtain a more realistic description of neutron stars,
one should also take into account the effects of a solid crust. In
{\em old neutron stars} a solid crust significantly affects the
size of the instability window\cite{BU00,LOU00,YL00,LU00} but has
been suggested to have no effect for the evolution of young
neutron stars. The size of the window is shown in Fig.~3, where we
discuss the influence of the $r$-mode instability in accreting old
neutron stars.

Strange stars, if they exist, might consist of the most stable
known form of matter, which is a mixture of strange, up and down
quarks. There are speculations that a considerable number of the
observed compact stars fall into this category. Both shear and bulk
viscosity are significantly different for strange stars, which has
a significant influence on the instability window\cite{Madsen98}.
Bulk viscosity is dominant for temperatures between
10$^8$-10$^{10}\,$K and as a result the instability window is
considerably smaller and moves to lower temperatures. As we will
show later, this has a significant effect for Low-Mass X-Ray
Binaries (LMXB)\cite{AJK02}. In addition a new instability window
appears for higher frequencies, see Fig.~7 in\cite{AK01}.

In specific EOS it is expected that the central parts of the neutron
star cores contain a significant number of {\em hyperons}.
Jones\cite{Jones01a,Jones01b} estimated that the bulk viscosity due to
non-leptonic weak interactions involving hyperons will be considerably
stronger. In a more detailed calculation Lindblom and Owen\cite{LO02}
have verified this result and have shown that the instability window
is considerably smaller. They suggested that for young neutron stars,
containing a considerable amount of hyperons in their cores, the r-mode
instability will be suppressed before it significantly affects the
angular momentum of the star.

\subsection{Evolution of nascent neutron stars}

Neutron stars are born after the gravitational collapse of the
neutron-star progenitor (massive star or accreting white dwarf).
Conservation of angular momentum suggests that they should rotate
almost at break-up speed, i.e.~near their Kepler limit. Moreover,
they should be quite hot with temperatures of the order of
10$^{11}\,$K. These nascent neutron stars cool down by neutrino
emission within a few seconds to temperatures below 10$^{10}\,$K
without considerable loss of angular momentum. According to Fig.~2
the star then enters the instability window and the amplitude of
the $r$-mode starts to grow exponentially. After a few minutes,
nonlinear effects take over and saturate the highly unstable mode.
When saturation occurs the star enters a nearly steady state, in
which the amplitude of the mode remains essentially unchanged and
the angular momentum of the star is radiated away while it
gradually cools down. After a year or so the angular velocity and
the temperature of the star are low enough that viscous damping
can start to dominate over the $r$-mode instability. Consequently,
the star moves out of the instability window and the phase of
angular momentum loss to gravitational radiation ends. The star
keeps cooling down slowly while its angular velocity remains
practically constant. In the case of the best known pulsar, the
Crab pulsar, this evolutionary scenario has provided some
impressive results. The Crab pulsar has an observed period of
$33\,$s, but 1000 years ago when it was born, it had an estimated
period of $19\,$s, which is exactly the period that the
evolutionary model predicts: Initially born with a period of
1-$2\,$ms, the Crab pulsar has radiated away about 95\% of its
angular momentum in the form of gravitational waves within a year,
thereby slowing down to a period of $\sim 19\,$s.

For young strange stars, it has been shown that the $r$-mode
driven spin-evolution is quite different from the neutron star
case\cite{AJK02}.  In a young strange star, the r-mode undergoes
short cycles of instability during the first few months, followed
by a quasi-adiabatic phase, where the $r$-mode remains at a small
and roughly constant amplitude for thousands of years. Another
important feature from the neutron star case is that the $r$-mode
in a strange star never grows to large amplitudes\cite{AJK02}.
These results suggest that the $r$-modes in young strange stars
emit a persistent gravitational-wave signal that should be
detectable with large-scale interferometers, given an observation
time of a few months\cite{AJK02}.

\subsection{Gravitational radiation emission}

The aforementioned 95\% loss of angular momentum due to the $r$-mode
instability is transformed into gravitational radiation, which is
emitted at a frequency range of 50-$1000\,$Hz. This large amount of
angular momentum loss in gravitational waves makes the rotational
instabilities one of the primary sources for the detection of
gravitational waves with laser interferometric detectors.

The frequency of the waves produced by an $l=2$, $m=-2$ $r$-mode
is $f_{\rm  gw}= 2\Omega /3\pi$ and the expected strain amplitude
is\cite{Owen98}
\begin{equation}
  h(t)=10^{-24} \ \alpha \ M_{1.4}\ R^3_{10}\left( {15 {\rm
        Mpc}\over D}\right) ,
\end{equation}
where $D$ is the distance from the source. The signal to noise ratio
for the above signal is
\begin{equation}
  \left({S\over N}\right)^2 = 2  \int^\infty_0 {|\tilde{h}(f)]^2df
    \over S_h(f)},
\end{equation}
where $\tilde{h}(f)$ is the Fourier transform of the signal and
$S_h(f)$ is the one-sided power spectral density of the detector
noise. If the saturation amplitude is about unity, newly formed
neutron stars out to about $10\,$Mpc should be detectable by LIGO
II with narrow banding. This suggests that the Virgo cluster is
likely to be out of reach\cite{BG98}. For the first phase of the
instability, when the signal grows up to a saturation point, there
are estimates predicting an upper signal to noise ratio of $S/N
<10$ for the LIGO II configuration\cite{OL02}.

\subsection{Accreting neutron stars}

The $r$-mode instability revived the interest in the gravitational
radiation producing mechanism suggested originally by Papaloizou
and Pringle\cite{PP78} and later also by Wagoner\cite{Wagoner}.
The mechanism applies to old neutron stars which are spun up by
accretion until they reach the instability window. For these stars
it was assumed that the loss of angular momentum via gravitational
wave should balance the accretion torque.

This scenario, initially suggested for $f$-modes, has been revived
for the $r$-modes\cite{AKSt99,Bildsten98}. It was soon
realized\cite{Levin99} that the viscous heating of the neutron
star will not allow the star to remain in the equilibrium state
suggested by Wagoner. Instead, the runway reheating will drive the
star further into the instability regime (path A-B in Fig.~3) and
spin it down (path B-C). This process will last for a few months
after which the star will follow the path (C-D-A), i.e.~it will
cool down and spin up again due to accretion. Eventually, it will
reach the instability window again and repeat this cycle.  This
quiet phase lasts for about 10$^6$ years, which means that an
accreting neutron star spends only a tiny fraction of its
accretion lifetime in the phase A-B-C, during which it emits
gravitational radiation. Still, Fig.~3 shows three remarkable
results\cite{AJKS00}. First the observed LMXB seem to follow the
pattern suggested here, second this mechanism sets a lower limit
for the rotation period of the recycled pulsars and third it
provides an explanation for the period clustering of the
millisecond pulsars. The point A in Fig.~3 corresponds to the
smallest possible period that the pulsar can achieve via
accretion. Remarkably, the theoretical value is $1.5\,$ms while
the fastest known pulsar rotates with a period of $1.56\,$ms.
Pulsars following this accretion scenario should have periods
bounded by the two points A and D, which correspond to 1.5 and
$6.5\,$ms and this is true for the majority of the known
millisecond pulsars.

\begin{figure}[t]
  \hbox to \hsize{\hfill \epsfysize=6cm \epsffile{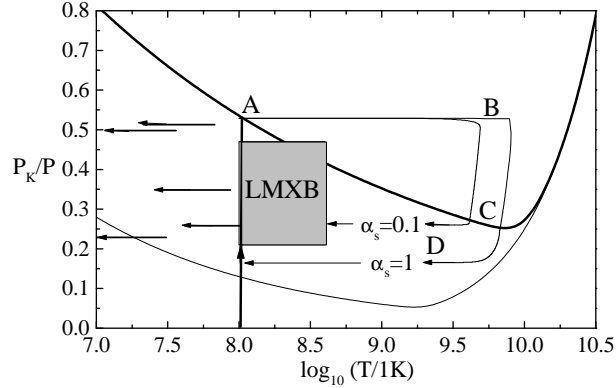}
    \hfill} \caption{ The $r$-mode instability window relevant for old
    neutron stars. We show results for the simplest (crust-free) model
    (thin solid line), as well as for a star with a crust (thick solid
    line).  We illustrate two typical $r$-mode cycles (for mode
    saturation amplitudes $\alpha_s=0.1$ and 1), resulting from
    thermo-gravitational runaway after the onset of instability.  For
    comparison with observational data, we indicate the possible range
    of spin-periods inferred from current LMXB data (shaded box) as
    well as the observed periods and estimated upper limits of the
    temperature of some of the most rapidly spinning millisecond
    pulsars (short arrows). }
  \label{evol3}
\end{figure}

Although Wagoner's original scenario is not realized for normal
neutron stars, it applies to strange stars. The instability window is
quite different here and it has been found that unstable $r$-modes
affect strange stars in a way that is quite distinct from the neutron
star case. For accreting strange stars, the onset of the $r$-mode
instability does not lead to the thermo-gravitational runaway that is
likely to occur in neutron stars\cite{AJK02}. Instead, the strange
star evolves towards a quasi-equilibrium state on a timescale of about
a year. The reason is that the instability window moves towards lower
temperatures and the corresponding point A is on the left side of the
window.  When the star becomes unstable, the runaway heating pushes
the star to higher temperatures i.e.~outside the window (ordinary
neutron stars are pushed further inside). The star becomes stable
again and soon cools down and ``hits" the instability line again and
the process continues for as long as the accretion lasts\cite{AJK02}.
This is shown in Fig.~4. The upper panel shows the time evolution of
the mode amplitude $\alpha$. After a few initial peaks, which are of
the order of $\alpha \sim 10^{-4}$, the amplitude settles to a
constant level of about $10^{-5}$.  During this ``stable'' phase the
mode heating remains also constant (middle panel) while the rotational
period of the star increases because of the continuous loss of angular
momentum in gravitational waves. This mechanism could explain the
clustering of spin-frequencies inferred from kHz QPO data in LMXB.
These results suggest that the $r$-modes in strange stars emit a
persistent gravitational-wave signal that should be detectable with
large-scale interferometers given an observation time of a few months.
If detected, these signals would provide unique evidence for the
existence of strange stars, which would put useful constraints on the
parameters of QCD.
\begin{figure}
  \hbox to \hsize{\hfill \epsfysize=5.5cm \epsffile{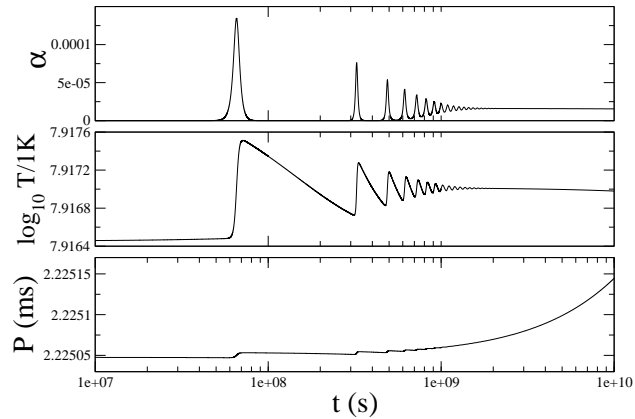}
    \hfill} \caption{The evolution of the $r$-mode amplitude (upper
    frame), the temperature (middle frame) and the spin period (bottom
    frame) corresponding to the first $10^{10}$~s following the
    initial onset of $r$-mode instability in an accreting strange
    star.  } \label{accrete}
\end{figure}

\subsection{Magnetic Fields}

Neutron stars usually have strong magnetic fields, which should
interact with a large amplitude pulsation mode. It has been suggested
that a magnetic field should have a strong influence on the evolution
of the $r$-mode instability and vice-versa.  Spruit\cite{Spruit}
suggested that the $r$-mode should generate some kind of differential
rotation.  In this case the magnetic field winds up until it reaches a
critical point where it becomes unstable due to buoyancy forces. When
this happens the star emits a considerable amount of electromagnetic
radiation and perhaps gamma rays. As a consequence the neutron star
will acquire an extremely strong magnetic field and become a magnetar.

Rezzolla, Lamb and Shapiro\cite{RLS00,RLS01a,RLS01b} suggested
that the magnetic field might actually prevent the $r$-mode
instability at all. This is possible if the non-linear evolution
of the $r$-modes winds up the magnetic field of the neutron star
in such a way that it drains energy away from the mode and thus
suppresses it entirely.  For magnetic fields of the order of
$B\geq 10^{10}\,$G, the rotation rate must be above 0.35$\,\Omega_K$
in order for the mode to survive, while for stronger magnetic
fields of the order of $B>10^{15}\,$G, the $r$-mode cannot grow at
all. Ho and Lai\cite{HL99} discussed the effect of magnetic
braking on the $r$-mode growth and showed that it is important
only for extremely strong magnetic fields $B \geq 10^{14}\,$G.  A
more systematic study of the effect of the magnetic field on the
$r$-modes was recently published by Morsink and Rezania\cite{MR01}.

\subsection{Nonlinear Calculations}

Many of the results that we presented till now were based on the
assumption that the $r$-modes reach an amplitude of order unity
and then saturate due to nonlinear processes. The way that the
nonlinear effects influence the saturation of a mode is not known
and there are questions which need to be answered. For example,
does the coupling of the $r$-mode to other modes allow the mode to
grow to unit amplitude? Does the growth of the mode influence an
initially uniformly rotating star in such a way that it becomes
differentially rotating? We have seen that this is an important
assumption for the influence of the magnetic field.

Nonlinear studies of pulsation modes suffer from the immense
computer power needed and both approaches that we will discuss are
limited by the computer resources. Stergioulas and Font\cite{SF01}
evolve a model of a rapidly rotating relativistic neutron star
using the nonlinear hydrodynamical equations without evolving the
spacetime (Cowling approximation). In this way one cannot see the
actual growth of the mode, but by evolving a sequence of perturbed
configurations with increasing initial mode amplitude $\alpha$,
one can study at least the energy leak from the $r$-mode into
other modes on a dynamical timescale. Though some other modes have
been excited, no mode coupling has been observed at amplitudes of
order unity at the resolution used in these simulations.

These results were verified by another nonlinear calculation by
Lindblom, Tohline and Vallisneri\cite{LTV01,Valli02}. They worked in
Newtonian theory and used a post-Newtonian radiation-reaction force to
drive the $r$-mode unstable. In order to shorten the growth time, they
increased the driving radiation reaction force by a factor of 4500.
In this way, they could observe the growth of the mode from an
amplitude $\alpha=0.1$ to $\alpha=2$. These results verified earlier
approximate calculations\cite{Owen98} of the loss of angular momentum
of the star.  They have not observed significant mode coupling either
but at an amplitude $a\sim 3.4$ they found nonlinear saturation, which
they argue was due to the creation of shocks associated with the
breaking of surface waves on the star.  In both calculations it was
observed that the $r$-modes generated differential rotation.

In two recent papers, Schenk at al.\cite{cornel1} and
Morsink\cite{Morsink02} developed a formalism to study the nonlinear
interaction of modes in rotating Newtonian stars. Subsequently they
used this formalism to study the saturation amplitude of the $r$-mode
instability\cite{cornel2}. Their results suggest that {\em the
  saturation amplitude is extremely small and never reaches amplitudes
  of order unity}. Instead they showed that the mode saturates at an
amplitude which is about four orders of magnitude smaller than that
suggested by Lindblom et al.\cite{LTV01}. This obvious disagreement
maybe a consequence of the limited resolution in the nonlinear time
evolutions\cite{SF01,LTV01}.

Recent estimations by Wagoner\cite{wagoner02} suggest, that even
if one takes into account the assumption that the r-mode amplitude
will never reach values of the order 1, the r-mode instability in
accreting neutron stars will still be a very good source for
gravitational waves. This result is similar to the one presented
earlier by Andersson et al.\cite{AJK02} for accreting strange
stars.

\section{Instabilities of rotating stars in GR}

The theory of non-radial perturbations of relativistic stars has
been a field of intensive study for more than three decades,
beginning with the pioneering paper of Thorne and
Campolattaro\cite{Thorne67}.

These authors focused on perturbations of non-rotating stars while
Hartle\cite{Hartle67} laid the foundations for computing rotating
relativistic stellar models. He also devised a way of modelling slowly
rotating stars.  This slow-rotation approximation is still widely used
since the problem becomes one dimensional and therefore much simpler
than the two dimensional case of rapidly rotating and strongly
deformed stars.

Investigating the axisymmetric perturbations, Chandrasekhar and
Ferrari\cite{CF91a} showed how rotation induces coupling of the polar
and axial modes, which are decoupled in the non-rotating case. Soon
after, Kojima\cite{Koj92} presented the first complete derivation of
the coupled polar and axial perturbation equations. Rotation does not
only couple the polar and axial modes, it also removes the
$(2l+1)$-fold degeneracy of a mode with angular index $l$ in a
non-rotating star.  To first order, this Zeeman-like frequency
splitting is linear in the rotation parameter.
Kojima\cite{Koj93a,Koj93b} computed the splitting for the relativistic
$f$-mode for various sequences of polytropic models. Yoshida and
Kojima\cite{YK97} compared the frequencies and their rotational
corrections for $f$ and $p$-modes in the relativistic Cowling
approximation with the fully relativistic perturbation calculation.
They concluded that the Cowling approximation is remarkably good and
becomes better the more relativistic the stellar models are.

The ``run'' on the $r$-modes started, when Andersson\cite{Nils98}
found that they are CFS-unstable for any rotation rate. In a first
attempt to compute the growth time within a fully relativistic
framework, he used the axial equation derived by
Kojima\cite{Koj92}, without the coupling to the polar equations.
Although he found the right sign of the imaginary part of the
complex $r$-mode frequencies, he was not able to recover the
$\Omega^{-(2l+2)}$ behavior (see eqn(19)). Ruoff and
Kokkotas\cite{RK02} pointed out that Kojima's equation is not
adequate for computing $r$-modes, for it is derived by
transforming the original axial field equations, which are first
order in time (or in the eigenfrequency $\sigma$), into a second
order wave equation. In these manipulations, the eigenvalue
$\sigma$ appears quadratically, and one obtains second order
rotational corrections, which, however, were discarded. Since the
$r$-mode frequency $\sigma$ is proportional to the rotation rate
$\Omega$, it is clear that one has to keep the $\Omega^2$ terms in
the wave equation.

Taking care of the proportionality between $\sigma$ and $\Omega$,
Kojima\cite{Koj98} derived an approximate equation in the so-called
low-frequency approximation, where the gravitational radiation
reaction is ignored.  This equation is now an eigenvalue equation,
which is linear in $\sigma$ and contains only rotational correction
terms linear in $\Omega$. Kojima immediately realized that the
relativistic treatment introduced a new effect, not present in the
Newtonian case, namely the appearance of a continuous spectrum, arising
from the relativistic frame dragging, which introduces an effective
$r$-dependent oscillation frequency. The presence of this continuous
spectrum in Kojima's equation has been proven in a mathematically
rigorous way by Beyer and Kokkotas\cite{BK99}.

For quite a while, it was not clear whether this continuous spectrum
was merely an artefact of the low-frequency approximation, which
neglects the radiation reaction, or whether it occurred because of the
neglect of the coupling to the polar equations, or a feature of the
slow-rotation approximation in general. Lockitch, Andersson and
Friedman\cite{LAF01} argued that Kojima's equation should only be
valid for non-barotropic stars, and showed that for uniform density
models, one can find regular mode solutions, representing the
relativistic $r$-modes.

Ruoff and Kokkotas\cite{RK01} and Yoshida\cite{Y01} extended this
study to polytropic models and found the somewhat surprising results
that Kojima's equations yields $r$-mode solution only for a very
restricted class of polytropic models, namely for those which small
polytropic index $n$. For large $n$, no regular solutions could be
found. Going one step further Ruoff and Kokkotas\cite{RK02} and Yoshida
and Futamase\cite{YF01} included the radiation reaction, but still
found that the basic properties of Kojima's equations do not change
and its singular structure due to the continuous spectrum cannot be
avoided.

Kojima and Hosonuma\cite{KH00} showed that if third-order rotational
effects are added to the original Kojima's equation, it becomes a
fourth order ordinary differential equation, which has no singular
properties if the Schwarzschild discriminant associated with the
buoyant force does not vanish inside the star. By solving a simplified
toy version of the extended Kojima's equation, Lockitch and
Andersson\cite{LA01} showed that the previously singular solutions can
be regularized when the additional fourth-order term is taken into
account.

Yoshida and Lee\cite{YL02} and Ruoff, Stavridis and
Kokkotas\cite{RSK02b} considered the fully coupled system in the
Cowling approximation by including only first order corrections. The
difference to the previous investigations is that the computation of
the mode frequencies and eigenfunctions is not done in a perturbative
method, in which $\Omega$ is regarded as a small expansion parameter.
Instead the mode is a solution of the complete coupled system, with
$l$ ranging from $l=m$ to infinity. For the numerical computation,
this system has to be truncated at some cut-off value $l_{\rm max}$
which has to be chosen such that the eigenfrequencies and
eigenfunctions are well converged as $l_{\rm max}$ is increased.

Yoshida and Lee\cite{YL02} focused on non-barotropic models taken from
McDermott, van Horn and Hansen\cite{MHH88}, which have a finite
$r$-dependent temperature. For all models under consideration, they
found regular $r$-mode solutions, and all of them were lying in the
regime of the continuous spectrum, which was forbidden from Kojima's
equation.

Ruoff et al.\cite{RSK02b} considered barotropic as well as
non-barotropic polytropic stellar models and demonstrated that in both
cases, one can always find $r$-modes. They discussed how the
truncation of the coupled system of equations at some cut-off value
$l_{\rm max}$ affects the continuous spectrum, which for the coupled
equations consists not only of a single band, but of several either
disconnected or overlapping patches. They found that some of the
inertial modes, in particular the $r$-mode, can actually exist inside
the continuous spectrum. However, for highly relativistic stellar
models, the continuous spectrum can still make some of the inertial
modes, which exist in less relativistic stars, disappear.

In the remainder of this review, we will outline the various
approximations and the results they are leading to. We will use
geometrical units with $G=c=1$.

\subsection{The slow-rotation formalism}

The metric describing a slowly rotating neutron star in spherical
coordinates ($t$, $r$, $\theta$, $\phi$) is
\begin{eqnarray}\label{metric}
  g_{\mu\nu} &=& \(
  \begin{array}{cccc}
    -e^{2\nu} & 0 & 0 & -\omega r^2 \sin^2\theta\\
    0 & e^{2\lam} & 0 & 0\\
    0 & 0 & r^2 & 0\\
    -\omega r^2 \sin^2\theta & 0 & 0 & r^2\sin^2\theta\\
  \end{array}\),
\end{eqnarray}
where $\nu$, $\lam$ and the ``frame dragging'' $\omega$ are
functions of the radial coordinate $r$ only. With the neutron star matter
described by a perfect fluid with pressure $p$, energy density
$\eps$, and 4-velocity
\begin{equation}
  U^\mu = e^{-\nu}\(1, 0, 0, \Omega\),
\end{equation}
the Einstein equations together with an equation of state $p=p(\eps)$
yield the well known TOV equations plus an extra equation for the
frame dragging, which to linear order is given by
\begin{equation}\label{drag}
  \varpi'' - \(4\pi re^{2\lam}(p + \eps) - \frac{4}{r}\)\varpi'
  - 16\pi e^{2\lam}\(p + \eps\)\varpi = 0,
\end{equation}
where
\begin{equation}
  \varpi := \Omega - \omega
\end{equation}
represents the angular velocity of the fluid relative to the local
inertial frame.

We assume the oscillations to be adiabatic, so that the relation
between the Eulerian pressure perturbation $\delta p$ and energy
density perturbation $\delta\eps$ is given by
\begin{equation}
  \label{adcond}
  \delta p = \frac{\Gamma_1p}{p + \eps}\delta\eps
  + p'\xi^r\(\frac{\Gamma_1}{\Gamma} - 1\),
\end{equation}
where $\Gamma_1$ represents the adiabatic index of the perturbed
configuration, $\Gamma$ is the background adiabatic index
\begin{equation}
  \Gamma = \frac{p + \eps}{p}\frac{p'}{\eps'},
\end{equation}
and $\xi^r$ is the radial component of the fluid displacement vector
$\xi^\mu$. If we write $\Gamma_1$ as
\begin{equation}
  \Gamma_1 = \frac{p + \eps}{p}\(\frac{\d p}{\d\eps}\)_s,
\end{equation}
we can write relation (\ref{adcond}) as
\begin{equation}
  \delta p = \(\frac{dp}{d\eps}\)_s\delta\eps - \xi^r{\cal A}_s
\end{equation}
with the Schwarzschild discriminant ${\cal A}_s$ given by equation (1).

The metric perturbations $h_{\mu\nu}$ are functions of all four
spacetime variables
\begin{equation}
  h_{\mu\nu} = h_{\mu\nu}(t,r,\theta,\phi)
\end{equation}
To eliminate the angular dependence, one expands the perturbations
into tensorial harmonics, which are a generalization of the well-known
spherical harmonics $Y_{lm} = Y_{lm}(\theta,\phi)$. One thus obtains a
set of equations for the coefficients, which only depend on $t$ and
$r$ and the parameters $l$ and $m$. In the non-rotating case, these
equations are independent of $m$ and decoupled with respect to $l$.
Furthermore, they fall into two independent sets, which show opposite
behavior under parity transformation. The {\it polar} or {\it even}
parity perturbations transform as $(-1)^l$, whereas the {\it axial} or
{\it even} parity perturbations as $(-1)^{l+1}$. Rotation induces a
coupling between the different sets, through first order rotational
terms, which couple the set of polar/axial equations of order $l$ to
the set of axial/polar equations of order $l\pm1$. Furthermore, the
equations now depend explicitly on $m$.

In deriving the perturbation equations, one has to make use of the
well-known gauge freedom. In the non-rotating case, the commonly used
gauge is the Regge-Wheeler gauge. However, in the rotating case, it is
more convenient to use a different gauge, the BCL gauge, which was
first introduced by Battiston, Cazzola and Lucaroni\cite{BCL71}. A
convenient way to set up this gauge and to derive the perturbation
equations for slowly rotating relativistic stars is to use the
ADM-formalism\cite{ADM62,RSK02a}. The ADM formalism provides a
natural framework to derive the perturbation equations for rotating
stars, as it clearly separates the equations into dynamical evolution
equations and the set of time independent constraints. As a
consequence the equations do not contain any mixed time and space
derivatives and in the BCL gauge, they do not contain any second
order derivatives at all.  This makes them particular suitable for the
numerical time evolution, and they can also be easily cast into an
eigenvalue problem.

In the ADM formalism, the gauge is chosen by giving prescriptions for
the lapse function $\alpha$ and the shift vector $\beta^i$. As we are
dealing with perturbations, $\alpha$ and $\beta^i$ are the
perturbations of lapse and shift. In the BCL-gauge, one sets $\alpha =
0$ and $\beta^i$ is chosen such that the evolution equations for the
angular perturbations $h_{\theta\theta}$, $h_{\theta\phi}$ and
$h_{\phi\phi}$ become
\begin{equation}\label{dthab}
  \d_th_{ab} = 0, \quad a,b = \{\theta,\phi\}.
\end{equation}
With $h_{ab} = 0$, the expansion of the metric perturbation is
\begin{eqnarray}
  \alpha &=& 0,\\
  \beta_i^{polar} &=& \sum_{l,m}{\(e^{2\lam}S_2^{lm},\,
    V_1^{lm}\d_\theta,\,V_1^{lm}\d_\phi\)Y_{lm}},\\
  h_{ij}^{polar} &=& \sum_{l,m}{\(\begin{array}{ccc}
    e^{2\lam}S_3^{lm} & V_3^{lm}\d_\theta & V_3^{lm}\d_\phi\\
    \star & 0 & 0\\
    \star & 0 & 0
  \end{array}\)Y_{lm}},\label{p_metric}
\end{eqnarray}
for the polar part and
\begin{eqnarray}
  \beta_i^{axial} &=& \sum_{l,m}{\(0, -V_2^{lm}\frac{\d_\phi}{\sin\theta},\,
    V_2^{lm}\sin\theta\d_\theta\) Y_{lm}},\\
  h_{ij}^{axial} &=& \sum_{l,m}{\(\begin{array}{ccc}
    0 & \displaystyle{-V_4^{lm}\frac{\d_\phi}{\sin\theta}}
    & V_4^{lm} \sin\theta\d_\theta\\
    \star & 0 & 0\\
    \star & 0 & 0
  \end{array}\)Y_{lm}}.
\end{eqnarray}
for the axial part. The asterisks stand for symmetric components. The
perturbations of the extrinsic curvature are all non-zero and denoted
by $K_1$ through $K_6$. Of these, $K_3$ and $K_6$ represent the axial
perturbations. Finally, the fluid perturbations are decomposed
according to
\begin{eqnarray}
  \delta u_i^{polar} &=& -e^\nu \sum_{l,m}{
    \(u_1^{lm}, u_2^{lm}\d_\theta, u_2^{lm}\d_\phi\)Y_{lm}},\\
  \delta u_i^{axial} &=& -e^\nu \sum_{l,m}{
    \(0, -u_3^{lm}\frac{\d_\phi}{\sin\theta},
    u_3^{lm}\sin\theta\d_\theta\)Y_{lm}},\\
  \delta \eps &=& \sum_{l,m}{\rho^{lm} Y_{lm}},\quad
  \delta p \;=\; \(p + \eps\)\sum_{l,m}{H^{lm} Y_{lm}},\\
  \xi^r &=& \bigg[\nu'\(1 - \frac{\Gamma_1}{\Gamma}\)\bigg]^{-1}
  \sum_{l,m}{\xi^{lm} Y_{lm}}.
\end{eqnarray}
From Eq.~(\ref{adcond}), we obtain the relation
\begin{equation}
  \rho^{lm} = \frac{\(p + \eps\)^2}{\Gamma_1p}\(H^{lm} - \xi^{lm}\).
\end{equation}
The requirement (\ref{dthab}) leads to the following three conditions
for the shift components
\begin{eqnarray}
  S_2^{lm} &=& \half K_4^{lm} - \omega e^{-2\lam}\(\I mV_3^{lm}
  + \cLa V_4^{lm}\),\\
  V_1^{lm} &=& K_5^{lm},\\
  V_2^{lm} &=& K_6^{lm},
\end{eqnarray}
where the operator $\cLa$ is defined by its action on a perturbation
variable $P^{lm}$
\begin{eqnarray}
  \cLa P^{lm} &=& (l-1)Q_{lm}P^{l-1m} - (l+2)Q_{l+1m}P^{l+1m},
\end{eqnarray}
with
\begin{eqnarray}
  Q_{lm} &:=& \sqrt{\frac{(l-m)(l+m)}{(2l-1)(2l+1)}}.
\end{eqnarray}
These requirements completely specify the gauge, and we obtain three
evolution equations for the metric perturbations $S_3$, $V_3$ and
$V_4$, as well as for the six extrinsic curvature components $K_1$ to
$K_6$. From the perturbed energy-momentum conservation
$\delta\(T^{\mu\nu}_{\phantom{\mu\nu};\mu}\) = 0$, we obtain the
evolution equations for the enthalpy pertubations $H$ and the velocity
perturbations $u_i$. In the non-barotropic case, we can derive the
evolution equation for $\xi$ from the relation (\ref{adcond}). In
addition to the evolution equations, the Einstein equations yield four
constraint equations which do not contain any time derivatives. The
complete set of equations is rather lengthy and can be found
in\cite{RSK02a}.

\subsection{Axial Equations}

We will first focus on the purely axial case.  Neglecting the
coupling to the polar equations, we have four evolution equations
for the metric variable $V_4$, the two axial components $K_3$ and
$K_6$ and the axial component of the 4-velocity $u_3$ (in the
following we omit the indices $l$ and $m$)
\begin{eqnarray}
  \label{V4}
  \(\df{}{t} + \I m\omega\)V_4 &=& e^{2\nu-2\lam}
  \left[K_6' + \(\nu' - \lam' - \frac{2}{r}\)K_6 - e^{2\lam}K_3\right]\;,\\
  \label{K3}
  \(\df{}{t} + \I m\omega\)K_3 &=& \frac{l(l+1) -2}{r^2}V_4
  + \frac{2\I m}{l(l+1)}\omega'e^{-2\lam}K_6\;,\\
  \label{K6}
  \(\df{}{t} + \I m\omega\)K_6 &=& V_4' - \frac{\I mr^2}{l(l+1)}
  \left[\omega'K_3 - 16\pi(\Omega - \omega)(p + \eps)u_3\right]\;,\\
  \label{u3}
  \(\df{}{t} + \I m\Omega\)u_3 &=& \frac{2\I m(\Omega-\omega)}{l(l+1)}
  \(u_3 - K_6\)\;,
\end{eqnarray}
together with the momentum constraint equation
\begin{equation}\label{MC}
  16\pi(p + \eps)u_3 = K_3' + \frac{2}{r}K_3
  - \frac{l(l+1) - 2}{r^2}K_6
  - \frac{2\I m\omega'}{l(l+1)}e^{-2\nu}V_4\;.
\end{equation}
In the literature, the two axial metric components are normally
denoted by $h_0$ and $h_1$. In the above equations, $V_4$ is
proportional to $h_1$ and $K_6$ is proportional to $h_0$. In the
non-rotating case $\dot u_3 = 0$, and the remaining equations
(\ref{V4}) -- (\ref{K6}) are equivalent to a single wave equation for
$V_4$, which in the exterior reduces to the well-known Regge-Wheeler
equation. For neutron stars, these equations describe the axial
$w$-modes, which have been first studied by Chandrasekhar and
Ferrari\cite{CF91b} for ultra-relativistic models, where they become
long lived ``trapped modes'', and by Kokkotas\cite{K94} for less
compact models.

In the rotating case, fluid motion is possible, and a new family
of modes is expected to appear, the unstable $r$-modes. For
ultra-compact and rapidly rotating stars, an ergoregion can
appear, which is able to drive some of the $w$-modes unstable.
This was known only for scalar and electromagnetic perturbations
\cite{Friedman78b} and has been investigated in detail by Commins
and Schutz\cite{CS78} for scalar fields on a stellar background.
In the following we will discuss it for the gravitational
perturbations themselves.

\subsection{The $w$-mode instability}

For modes mainly associated with fluid motion, it is the rotation of
the star itself which is responsible for dragging a counter-rotating
mode forward. For $w$-modes, which are predominantly spacetime
oscillations it is not the stellar rotation itself, but the associated
frame dragging, that influences the behavior of the $w$-modes. In the
non-rotating case, the axial $w$-modes do not couple to any fluid
oscillations at all. If the star is set into rotation, this remains
only approximately true as the fluid motion is induced by
Eq.~(\ref{u3}), and in principle also through the coupling to the
polar equations.

But if one still regards the $w$-mode oscillations as being
essentially independent of the fluid, then it becomes clear that
the only way to make a $w$-mode unstable is to create an
ergosphere, i.e.~a region where no timelike counter-rotating
geodesics exist any more, or in other words, a region where the
dragging is so strong that any timelike backwards moving
trajectory gets dragged forward. This is associated with a change
in the sign of $g_{tt}$.

To test the existence of an ergosphere we write $g_{tt}$
as\cite{CS78}
\begin{equation}
  \label{eq:gtt}
  g_{tt} = -e^{2\nu} + \omega^2r^2\sin^2\theta
\end{equation}
and restrict ourselves to the equatorial plane $\theta = \pi/2$.
An ergosphere starts to appear whenever $g_{tt} > 0$, or
equivalently when
\begin{equation}
  \label{eq:ergo}
  e^{\nu} < r\omega\;.
\end{equation}
Models allowing ergospheres have been constructed in the case of
rapidly rotating uniform density models by Butterworth and
Ipser\cite{BI76} and for toroidal polytropic configurations by
Komatsu, Eriguchi and Hachisu\cite{KEH89}. It is clear that for
any realistic equation of state, it is very unlikely to construct
stable stellar models which would exhibit an ergosphere,
nevertheless recent results suggest the existense of ultra compact
quark stars\cite{Drake02}. Therefore, according to the present
observational data and the known types of EOS, one would not
expect the ergosphere instability of the $w$-modes to be of any
significant astrophysical relevance. The compactness needed for a
star to develop ergosphere is extremely high i.e.~$R/M\sim
2.26$-$2.4$. An interesting feature of this instability is that it
can set in even for quite low rotation rates. For example for a
star with $R/M=2.26$ the first $w$-mode becomes unstable at a
rotation rate of $0.19\,\Omega_K$. For $R/M=2.4$ this occurs for
$0.83\,\Omega_K$. Given that a star is becoming unstable due to
the existence of an ergosphere then there is no direct dissipation
due to shear or bulk viscosity, since the $w$-modes practically do
not couple to the fluid motions! This is a intriguing result since
it suggests that the instability will grow and the star will
radiate away in gravitational waves all its available angular
momentum before it becomes stable again\cite{KRA02}.

\begin{figure}[t]
\begin{center}
\epsfxsize=8cm \epsfbox{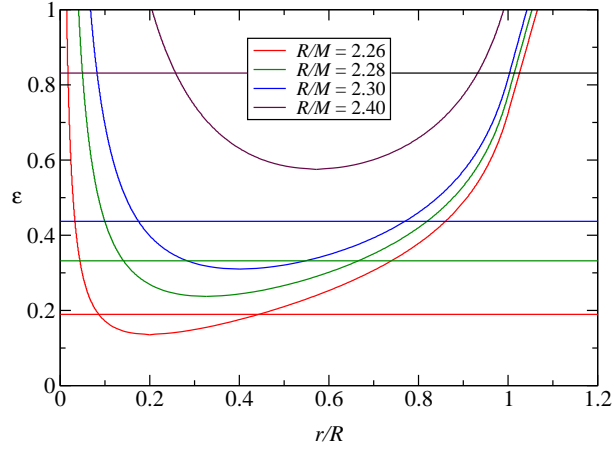} \vspace*{5mm}
\caption{\label{fig1}Appearance of the ergoregion as a function
  of the rotation rate $\veps=\Omega/\Omega_K$. For the stellar model with $R/M = 2.4$,
  it first appears at $r = 0.2R$ for $\veps = 0.15$ and expands as
  $\veps$ is increased. The vertical lines represent the rotation
  rates, at which the first $w$-mode of the respective models becomes
  unstable.}
\end{center}
\end{figure}

\begin{figure}[t]
\begin{center}
\epsfxsize=8cm \epsfbox{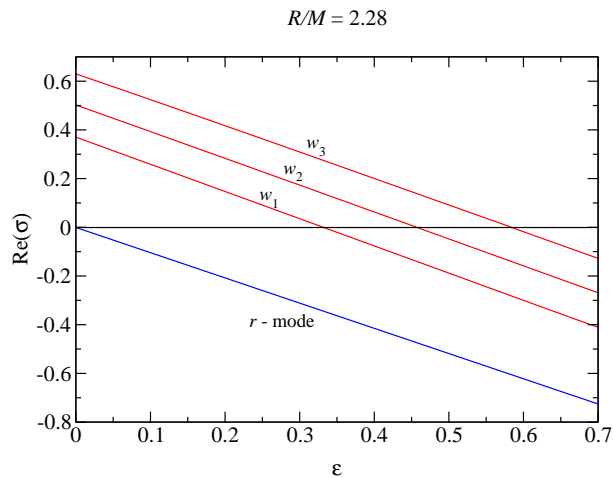} \vspace*{5mm}
\caption{\label{fig2}The real part of the frequency of the three
  $w$-mode together with the $r$-mode of the stellar model with $R/M =
  2.28$. The $w_1$-mode crosses the zero line at $\veps = 0.32$.}
\end{center}
\end{figure}

\subsection{The $r$-mode instability}

In contrast to the $w$-modes, which are predominantly spacetime
oscillations, the $r$-modes are characterized by strong axial fluid
currents. Even though, it is only through the metric perturbations,
that the $r$-modes are driven unstable, it is instructive to study
their features in various approximations, where some or all metric
perturbations are neglected.

The most radical way is to neglect all metric perturbations. For modes
with small damping, such as the polar $p$-modes for example, this
so-called relativistic Cowling approximation yields quite accurate
results. In Newtonian theory, it also yields very reliable results for
the $r$-modes. Let us therefore apply it to the relativistic case as
well. Neglecting all metric perturbation in our set of axial evolution
equations (\ref{V4}) -- (\ref{u3}), the only remaining equation is that
for the fluid variable $u_3$, which can be written as
\begin{equation}\label{Cowl}
  \df{}{t}u_3 = -im\(\Omega - {\frac{2\varpi}{l(l+1)}}\)u_3\;.
\end{equation}
From this equation we can immediately deduce that the fluid layers are
decoupled from each other, which means that each layer has its own
real oscillation frequency given by
\begin{equation}\label{freqCowl}
  \sigma = -m\(\Omega - \frac{2\varpi}{l(l+1)}\)\;.
\end{equation}
The whole frequency range is determined by the values of $\varpi$ at
the center and at the stellar surface, and is the larger the more
relativistic the stellar model is. In the Newtonian limit $(\varpi
\rightarrow \Omega)$, Eq.~(\ref{freqCowl}) yields the well known
$r$-mode frequency
\begin{equation}\label{freqNewt}
  \sigma_{\rm N} = -m\Omega\(1 - \frac{2}{l(l+1)}\)\;.
\end{equation}
In the relativistic case, the presence of the frame dragging $\omega$
destroys the occurrence of a single mode frequency and gives rise to a
continuous spectrum, at least to this order of approximation.

This somewhat peculiar result might suggest that the Cowling
approximation fails in the relativistic case, and we should rather
keep some of the metric perturbations. To assess which ones to keep,
we consider the non-rotating case. There $\dot u_3 = 0$, i.e.~we have
no oscillatory motion.  We can have, however, have a stationary fluid
current with a nonzero associated metric perturbation, which we can
easily find by setting in our evolution equations (\ref{V4}) --
(\ref{u3}) all the time derivatives to zero (together with the
rotational terms). From equations (\ref{K3}) and (\ref{K6}), we
immediately see that $V_4$ has to vanish and from Eq.~(\ref{V4}) we
obtain
\begin{eqnarray}
  \label{K6_0}
  K_6' + \(\nu' - \lam' - \frac{2}{r}\)K_6 - e^{2\lam}K_3 &=& 0\;.
\end{eqnarray}
Moreover, we have the constraint (\ref{MC}), which we can combine with
the above equation to yield a single equation for $K_6$. To this end,
we differentiate (\ref{K6_0}) with respect to $r$ and use the
constraint (\ref{MC}) to eliminate $K_3$. This gives us an elliptic
second order equation for $K_6$ with a source term containing the
fluid velocity $u_3$. Instead of using $K_6$, it is more convenient to
write this equation in terms of the variable $h_0 = e^{\nu-\lam} K_6$
\begin{equation}
  \label{eqh0}
  e^{-2\lam} h_0''-4\pi r(p + \eps)h_0' + \left[8\pi(p + \eps)
    + \frac{4M}{r^3} - \frac{l(l+1)}{r^2}\right]h_0
  = 16\pi e^{\nu-\lam}(p + \eps)u_3.
\end{equation}
Although this equation is derived under the assumption of stationary
perturbations, we can also use it for slowly rotating starts and
low-frequency oscillations. In this case we no longer assume $\dot u_3
= 0$, but we use the evolution equation (\ref{u3}) to govern the
time dependence of $u_3$.

It is obvious that the above equation (\ref{eqh0}) together with the
evolution equation (\ref{u3}) is qualitatively different from the
Cowling approximation (\ref{Cowl}). In the latter case, the fluid
layers are completely decoupled and each one oscillates with its own
frequency. With the inclusion of $h_0$, we have an elliptic
Poisson-like equation, with the fluid acting as a source. As $h_0$
couples back into the evolution equation (\ref{u3}), it connects
the fluid cells and might give rise to some coherent motion.

\begin{figure}[t]
\begin{center}
\vspace*{1cm} \leavevmode \epsfxsize=8cm \epsfbox{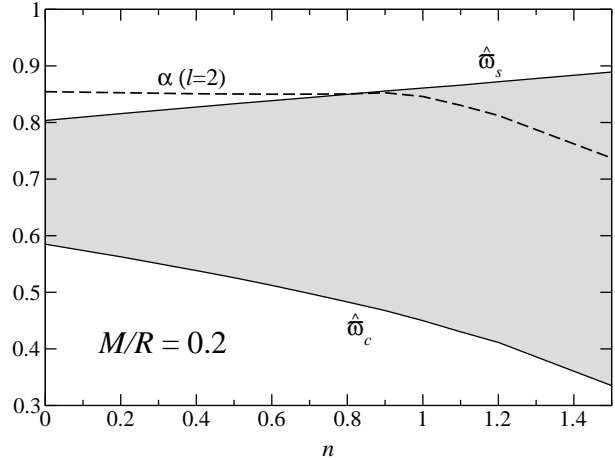}
\vspace*{5mm} \caption{\label{fig3}
  The boundaries of the continuous spectrum $\hat\varpi_c$ and
  $\hat\varpi_s$ together with the $r$-mode frequency $\alpha$ as a
  function of the polytropic index $n$ for stellar models with
  compactness of $M/R = 0.2$.  The case $n=0$ corresponds to a
  uniform density model.}
\end{center}
\end{figure}

To find out whether we can find mode solutions, we assume a harmonic
time dependence of $u_3$, i.e. $u_3 \sim \exp(\I\sigma t)$. With this
ansatz, we can combine equations (\ref{u3}) and (\ref{eqh0}) into a
single equation, which can be written as
\begin{eqnarray}\label{eveqn_a}
  &&\(\alpha - \hat\varpi\)
  \left[e^{-2\lam} h_0''
    -4\pi r(p + \eps)h_0' -\(8\pi(p + \eps) - \frac{4M}{r^3}
    + \frac{l(l+1)}{r^2}\)h_0\right]\non\\
  &&{}\qquad\qquad+ 16\pi(p + \eps)\alpha h_0 = 0\;,
\end{eqnarray}
where we have introduced the normalized frequency
\begin{equation}
  \alpha = \half l(l+1)\(1 + \frac{\sigma}{m\Omega}\)
\end{equation}
and
\begin{equation}
  \hat\varpi := \varpi/\Omega\;.
\end{equation}
The relation between $u_3$ and $h_0$ becomes
\begin{equation}\label{u3alg_a}
   u_3 = \frac{\hat\varpi}
  {\hat\varpi - \alpha}e^{\lam-\nu} h_0\;.
\end{equation}
Eq.~(\ref{eveqn_a}), which is Kojima's master equation\cite{Koj98}, is a
singular eigenvalue problem, because the factor $\alpha - \hat\varpi$
becomes zero if $\alpha$ falls into the range
\begin{equation}
  \hat\varpi_c < \alpha < \hat\varpi_s\;.
\end{equation}
In this case, the fluid perturbation diverges unless $h_0$ vanishes at
the singular point, which one can prove not to be the case\cite{RK01}.
If we translate the singular range for $\alpha$ back into the
corresponding range for the eigenfrequency $\sigma$, we see that it is
exactly the same as the continuous frequency spectrum we found in the
Cowling approximation given by Eq.~(\ref{freqCowl}).

One can show that mode solutions can only exist for $\alpha \le
1$\cite{LAF01}.  If $\alpha$ lies in the range $\hat\varpi_s < \alpha
\le 1$, the eigenvalue equation is no longer singular, as the singular
point now lies outside the star, where $u_3$ is zero. Only for such
values of $\alpha$, regular mode solutions can be found.

For uniform density models one can indeed find regular mode
solutions\cite{LAF01}. However for polytropic equations of state, this
is only possible for a very restricted range of stellar
models\cite{RK01,Y01}. The existence of regular mode solutions
strongly depends on the polytropic index $n$ and the compactness of
the stellar model under consideration. The general picture is that the
smaller $n$ is, the larger is the compactness range where one can find
physical mode solutions. For a given polytropic index $n$, one usually
finds physical mode solutions for models with a small $M/R$ ratio. As
the compactness is increased, i.e.~as the models become more
relativistic, the mode frequency $\alpha$ decreases and starts
approaching $\hat\varpi_s$.  Eventually it crosses this point and
migrates inside the range of the continuous spectrum, thus becoming
unphysical, and no $r$-mode exists any more. This is shown in
Fig.~\ref{fig3}.

It has been argued that the singular structure which is associated
with the continuous spectrum is an artefact of neglecting the
radiation reaction. Both in the Cowling and low-frequency
approximations, the eigenfrequency $\sigma$ is always real, whereas it
actually should be complex since the emission of gravitational waves
should either damp the modes or make them grow. If the frequency is
complex, the singular point is removed from the real axis, and the
differential equation is no longer singular.

\begin{figure}[t]
\begin{center}
\vspace*{1cm} \leavevmode \epsfxsize=8cm \epsfbox{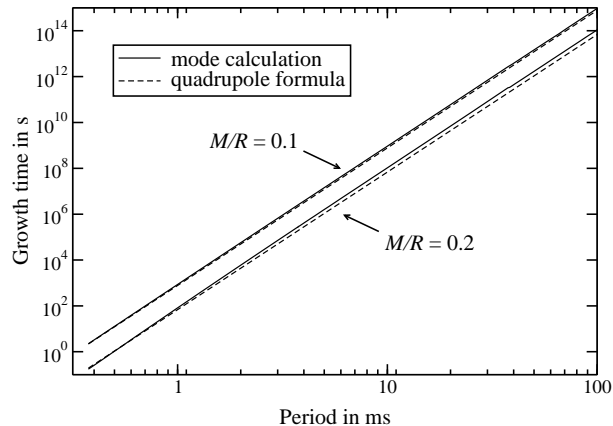}
\vspace*{5mm} \caption{\label{fig4}
  Double logarithmic plot of the growth time for uniform density
  models with compactness $M/R$ = 0.1, 0.2. Also included is the
  growth time deduced from formula (\ref{g_rate}), which agree well
  with the numerical results.}
\end{center}
\end{figure}

That the inclusion of the radiation reaction does not resolve this
issue has been shown by Ruoff and Kokkotas\cite{RK02} by solving the
full set of axial equations and by Yoshida and Futamase\cite{YF01} by
imposing a near-zone boundary condition for Kojima's equations instead
of the asymptotic flatness condition.

The conversion of the evolution system (\ref{V4}) -- (\ref{u3}) into a
time independent form is straightforward. Assuming a harmonic time
dependence $e^{i\sigma t}$, we can deduce the following two ordinary
differential equations
\begin{eqnarray}
  \label{V4ode}
  V_4' &=& \frac{\I mr^2}{l(l+1)}
  \(\omega'K_3 - 16\pi\varpi(p + \eps)u_3\)
  + \I\(\sigma + m\omega\)K_6\;,\\
  \label{K6ode}
  K_6' &=& e^{2\lam}K_3 - \(\nu' - \lam' - \frac{2}{r}\)K_6
  + \I e^{2\lam-2\nu}\(\sigma + m\omega\)V_4\;,
\end{eqnarray}
together with the two algebraic relations
\begin{eqnarray}
  \label{K3alg}
  K_3 &=& \frac{-\I}{\sigma + m\omega}\(\frac{l(l+1) -2}{r^2}V_4
  + \frac{2\I m\omega'}{l(l+1)}e^{-2\lam}K_6\)\;,\\
  \label{u3alg}
  u_3 &=& \frac{2m\varpi}
  {2m\varpi - l(l+1)(\sigma + m\Omega)}K_6\;.
\end{eqnarray}
Again, Eq.~(\ref{u3alg}) becomes singular if $\sigma$ is real and
falls in the range of the continuous spectrum. For real $\sigma$ the
denominator $\sigma + m\omega$ in Eq.~(\ref{K3alg}) can also vanish
and the equation becomes ill defined.  However, as $\sigma$ is
expected to be a complex frequency, both equations for $K_3$ and $u_3$
should remain regular.

For uniform density models, the low-frequency approximation always
yielded $r$-mode solutions. The inclusion of the radiation reaction
should then yield the associated growth times. For stellar models,
which are not too compact, the fully relativistic calculation can
indeed reproduce the Newtonian value given for $l=m=2$ by\cite{KS99}
\begin{equation}
  \label{g_rate}
  \tau_{\rm PN} = 22\(\frac{1.4M_\odot}{M}\)
  \(\frac{\mbox{10 km}}{R}\)^4\(\frac{P}{\mbox{1 ms}}\)^6\mbox{s}
\end{equation}
with a remarkably good agreement. For very compact models, however,
the growth times found in the fully relativistic case are in general
larger than suggested by formula (\ref{g_rate}). In Fig.~\ref{fig4}
we show the growth times for two uniform density models with $M/R =
0.1$ and 0.2 as function of the rotation period.

\begin{figure}[t]
\begin{center}
\epsfxsize=9cm \epsfbox{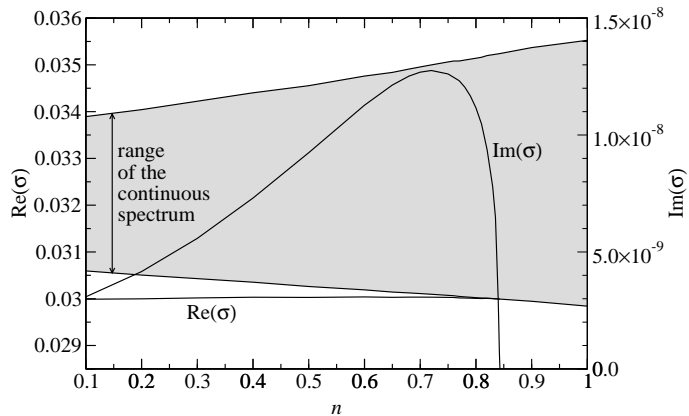} \vspace*{5mm}
\caption{\label{fig5}Behavior of the real and imaginary part of
the  $r$-mode frequency $\sigma$ as a function of the polytropic
index
  $n$ for models with $M/R = 0.2$ and rotation period of $1.0\,$ms.
  The scale on the left is for the real part, the scale on the right
  for the imaginary part of $\sigma$. For small $n$, the real part
  Re$(\sigma)$ lies outside the range of the continuous spectrum.  For
  increasing $n$, Re$(\sigma)$ approaches the lower boundary and the
  imaginary part Im$(\sigma)$ eventually starts to decrease.  For $n =
  0.843$, $\Re(\sigma)$ reaches the lower boundary of the continuous
  spectrum and Im$(\sigma)$ goes to zero. For larger $n$, no mode can
  be found any more.}
\end{center}
\end{figure}

When switching to polytropes, we still find the same pictures as in
the low-frequency approximation. As long as the mode is well outside
the continuous spectrum, we find a non-vanishing imaginary part, but
as soon as the mode approaches the continuous spectrum, the imaginary
part starts to decrease and drops to zero at the point where the real
part enters the continuous spectrum. This is depicted in
Fig.~\ref{fig5}, where the dependence of the complex eigenfrequency
$\sigma$ on the polytropic index $n$ is shown. For small $n$, the real
part of $\sigma$ lies well outside the continuous spectrum, but as $n$
increases, the continuous spectrum expands and approaches the real
part of $\sigma$. The imaginary part first grows with $n$, but as the
continuous spectrum expands and approaches the real part of $\sigma$,
it bends down and goes to zero very rapidly. For $n = 0.843$, is
becomes zero and for larger $n$, a regular mode no longer exists.

\subsection{Inclusion of the coupling}

The above investigations have shown that purely axial modes do not
necessarily exist in relativistic non-barotropic stars. So far we have
assumed that the coupling to the polar equations can be neglected.
However, as the results of the previous sections have shown, this
might lead to very unexpected results. To assess, whether the existence
of the $r$-modes indeed depends on the stellar model, one has to
go one step further and include the coupling to the polar equations.
As the complete set of equations is quite complicated, it is useful
to study the effect of the coupling within some further approximations.

Although the Cowling approximation did not give any useful results in
the purely axial case, it nevertheless yields the key to solving the
puzzling results of the previous sections.

The first order evolution equations in the Cowling approximation are
derived from the perturbed energy-momentum conservation law
$\delta\(T^{\mu\nu}_{\phantom{\mu\nu};\mu}\) = 0$, and read in the
non-barotropic case\cite{RSK02a}:
\begin{eqnarray}
  \label{dtHb}
  \(\d_t + \I m\Omega\)H &=& e^{2\nu-2\lam}\bigg\{C_s^2\bigg[u_1' +
  \(2\nu' - \lam' + \frac{2}{r}\)u_1 - e^{2\lam}\frac{l(l+1)}{r^2}u_2\non\\
  &&\qquad\qquad+ 2\I m\varpi e^{2\lam-2\nu}H\bigg] - \nu'u_1\bigg\},\\
  \label{dtu1b}
  \(\d_t + \I m\Omega\)u_1 &=& H'
  + \frac{p'}{\Gamma_1 p}\bigg[\(\frac{\Gamma_1}{\Gamma} - 1\)H + \xi\bigg]
  - B\(\I mu_2 + \cLa u_3\),\\
  \label{dtu2b}
  \(\d_t + \I m\Omega\)u_2 &=& H
  + \frac{2\varpi}{l(l+1)}\(\I mu_2 + \cLc u_3\)
  - \I m\frac{r^2e^{-2\lam}}{l(l+1)}Bu_1,\\
  \label{dtu3b}
  \(\d_t + \I m\Omega\)u_3 &=&
  \frac{2\varpi}{l(l+1)}\(\I m u_3 - \cLc u_2\)
  + \frac{r^2e^{-2\lam}}{l(l+1)}B\cLb u_1,\\
  \label{dtxib}
  \(\d_t + \I m\Omega\)\xi &=& \nu'\(\frac{\Gamma_1}{\Gamma} - 1\)
  e^{2\nu-2\lam}u_1,
\end{eqnarray}
with
\begin{eqnarray}
  B &=& \omega' + 2\varpi\(\nu' - \frac{1}{r}\),
\end{eqnarray}
and the coupling operators
\begin{eqnarray}
  \label{cLa}
  \cLa P^{lm} &=& (l-1)Q_{lm}P^{l-1m} - (l+2)Q_{l+1m}P^{l+1m},\\
  \label{cLb}
  \cLb P^{lm} &=& -(l+1)Q_{lm}P^{l-1m} + lQ_{l+1m}P^{l+1m},\\
  \label{cLc}
  \cLc P^{lm} &=& (l-1)(l+1)Q_{lm}P^{l-1m} + l(l+2)Q_{l+1m}P^{l+1m}.
\end{eqnarray}
As these operators couple equations of order $l$ to equations of
order $l\pm1$, the above set of equations represents a coupled system
with $l$ ranging from $l=m$ to $l=\infty$.

\subsection{The Continuous Spectrum}

By assuming a harmonic time dependence, one can easily transform the
above time dependent equations into an eigenvalue problem for the
frequency $\sigma$. Defining
\begin{eqnarray}
  u^l &=& \(\begin{array}{c}u_2^l\\u_3^l\end{array}\),\quad
  s^l \;=\; \(\begin{array}{c}H^l\\u_1^l\end{array}\),\quad
  l = m, \dots, \infty
\end{eqnarray}
the eigenvalue equations can write be written as a set of ODEs
\begin{eqnarray}\label{evsys}
  (s^l)' &=& As^l + B^l_-u^{l-1} + B^lu^l +  B^l_+u^{l+1}
  \quad l = m, \dots, \infty
\end{eqnarray}
together with the algebraic relations
\begin{eqnarray}
  \label{US}
  U^l_-u^{l-1} + U^l_\Sigma u^l + U^l_+u^{l+1} &=&
  S^l_-s^{l-1}+ S^ls^l+ S^l_+s^{l+1},\quad l = m, \dots, \infty,
\end{eqnarray}
where the letters $A$, $B$, $U$ and $S$ with their various indices
represent $2\times 2$ matrices. The algebraic relations can be viewed
as a matrix equation for two infinitely dimensional matrices {\sf U}
and {\sf S} acting on the vectors
\begin{eqnarray}
  u &=& \(u^m, u^{m+1}, u^{m+2}, \dots, u^l, \dots\)^{\rm T},\\
  s &=& \(s^m, s^{m+1}, s^{m+2}, \dots, s^l, \dots\)^{\rm T},
\end{eqnarray}
whose respective elements are the 2-vectors $u^l$ and $s^l$. Both {\sf
  U} and {\sf S} are tridiagonal block matrices, with each block given
by the above $2\times2$ matrices. Now we can write Eq.~(\ref{US}) as
\begin{eqnarray}
  \label{USb}
  \sum_{l'}{\mbox{\sf U}^{ll'}u^{l'}}
  &=& \sum_{l'}{\mbox{\sf S}^{ll'}s^{l'}}\;, \qquad l = m\dots\infty,
\end{eqnarray}
which can be solved for $u^l$ by multiplying both sides with ${\sf
  U}^{-1}$
\begin{eqnarray}
  \label{matrix}
  u^l &=& \sum_{l'l''}{\(\mbox{\sf U}^{-1}\)^{ll'}
    \mbox{\sf S}^{l'l''}s^{l''}}\;, \qquad l = m\dots\infty.
\end{eqnarray}
Note that the matrix {\sf U} is $r$-dependent since its elements
$U^l_\pm$ contain the function $\varpi$. For {\sf U} to be invertible
its determinant must not vanish. For certain values of $\sigma$,
however, this happens. In the Newtonian limit
$\varpi\rightarrow\Omega$, the matrix {\sf U} becomes independent of
$r$, and one can easily show that {\sf U} becomes singular only for a
discrete set of frequencies $\sigma$. These frequencies represent the
solutions of the homogeneous part ${\sf U}u = 0$ of Eq.~(\ref{USb}).
In the relativistic case, however, the zeroes of $\det {\sf U}$ depend
on the position inside the star, which means that each single
Newtonian frequency will be spread out into a continuous band of
frequencies, determined by the values of $\varpi$ at the center and at
the surface of the star, which we will denote by $\varpi_0$ and
$\varpi_R$, respectively. In order to numerically perform the
inversion and solve equation (\ref{matrix}), we have to truncate the
system at some value $l_{\rm max} \ge m$. To explicitly compute the
ranges of the resulting continuous spectrum we need the matrices
$U^l_-$, $U^l_+$ and $U^l_\Sigma$, which are given by
\begin{eqnarray}
  U^l_- &=& -2\hat\varpi\frac{l-1}{l}Q_{lm}\(
  \begin{array}{cc}0&1\\1&0\end{array}\),\\
  U^l_\Sigma &=& \Sigma
  \(\begin{array}{cc}1&0\\0&1\end{array}\),\\
  U^l_+ &=& -2\hat\varpi\frac{l+2}{l+1}Q_{l+1m}\(
  \begin{array}{cc}0&1\\1&0\end{array}\),
\end{eqnarray}
with
\begin{equation}
  \Sigma = \kappa - \frac{2m\varpi}{l(l+1)}
\end{equation}
and
\begin{equation}
  \kappa = \sigma + m\Omega.
\end{equation}
In the following examples, we choose $m=2$. Let us start with the
simplest case which is choosing $l_{\rm max} = 2$. Then ${\sf U} =
U^2_\Sigma $ and $\det{\sf U} = 0$ just yields
\begin{equation}
  \Sigma = 0,
\end{equation}
which gives exactly the same continuous spectrum as following from
Eq.~(\ref{freqCowl}). For $l_{\rm max} = 3$
\begin{equation}
  {\sf U} = \(\begin{array}{cc}U^2_\Sigma&U^2_+\\
    U^3_-&U^3_\Sigma\end{array}\),
\end{equation}
and $\det{\sf U} = 0$ yields the two solutions
\begin{eqnarray}
  \kappa &=& \(\half \pm \frac{\sqrt{105}}{14}\)\varpi.
\end{eqnarray}
Hence, the two ranges of the continuous spectrum are given by
\begin{eqnarray}
  1.2319\,\varpi_0\le\kappa\le 1.2319\,\varpi_R,
\end{eqnarray}
for the plus sign and
\begin{eqnarray}
  -0.2319\,\varpi_0\le\kappa\le -0.2319\,\varpi_R,
\end{eqnarray}
for the minus sign. For $l_{\rm max} = 4$, we obtain
\begin{equation}
  \frac{\kappa}{\varpi} = \left\{\begin{array}{c}1.4964\\0.4669\\-0.7633\\\end{array}.\right.
\end{equation}
Each increment of $l_{\rm max}$ leads to an additional range of the
continuous spectrum. Depending on the respective values of $\varpi_0$
and $\varpi_R$, the various ranges might as well overlap.  This
actually happens for the more relativistic stars since the variation
in $\varpi$ is much greater.

\begin{figure}[t]
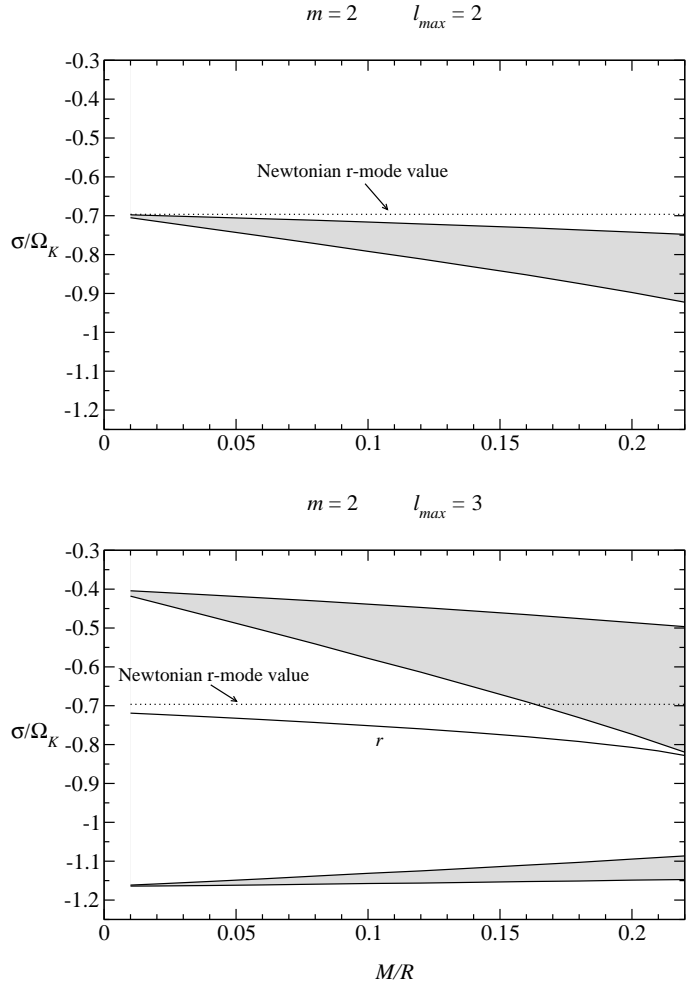

  \centering
  \epsfxsize=9cm
  \epsfbox{figure10a.eps}
  \epsfxsize=9cm
  \epsfbox{figure10b.eps}
  \caption{\label{fig6}The continuous spectrum and the $r$-mode for $m=2$
    as functions of $M/R$ for $l_{\rm max} = 2$ and $l_{\rm max} = 3$.
    For $l_{\rm max} = 2$, only the continuous spectrum exists. The
    relativistic $r$-mode appears for $l_{\rm max} = 3$ well outside
    the continuous spectrum as the latter has shifted away from its
    previous location for $l_{\rm max} = 2$. The dotted line represent
    the Newtonian $r$-mode value from Eq.~(\ref{freqNewt}).}
\end{figure}
\begin{figure}[t]
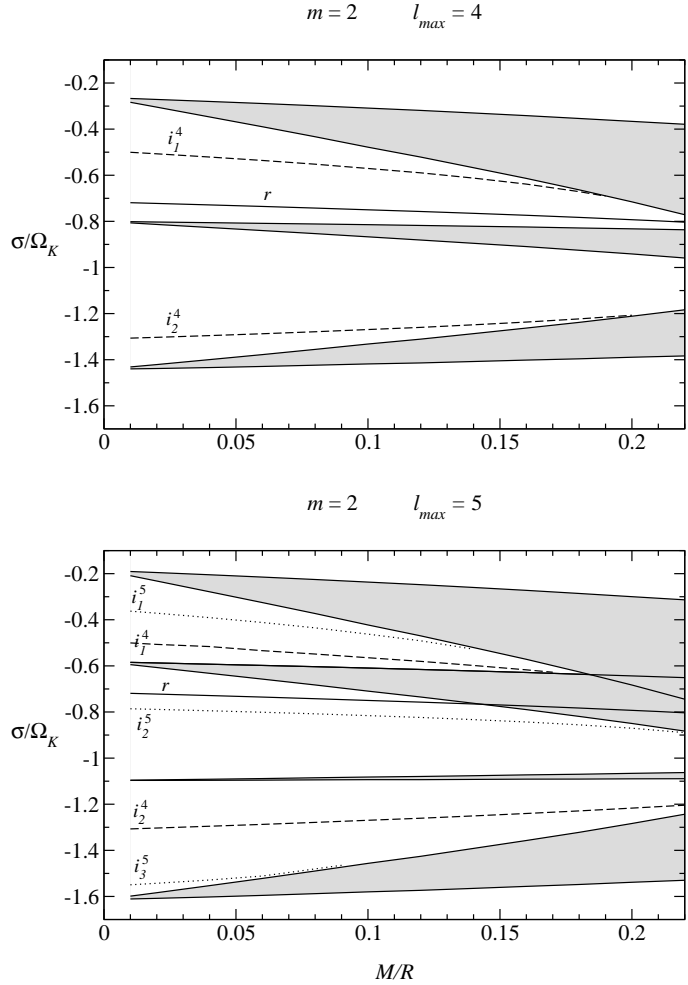

  \centering
  \epsfxsize=9cm
  \epsfbox{figure11a.eps}
  \epsfxsize=9cm
  \epsfbox{figure11b.eps}
  \caption{\label{fig7}Same as in Fig.~10 for $m=2$. In addition we
    have plotted the Newtonian value for the $r$-mode according to
    Formula (\ref{freqNewt}). The relativistic $r$-mode appears for
    $l_{\rm max} = 3$ well outside the continuous spectrum as the
    latter has shifted away from its previous location for $l_{\rm
      max} = 2$.  Only the $r$-mode and the $i_2^4$ and $i_2^5$ modes
    can exist for the whole compactness range.}
\end{figure}

In the Newtonian limit $\varpi \rightarrow \Omega$, the continuous
spectrum shrinks down to discrete values. Interestingly, these values
$\kappa/\Omega$, following from $\det{\sf U} = 0$, are exactly the
eigenvalues $\kappa_0$ of the (generalized) $r$-modes of the Maclaurin
spheroids in the limit of low angular velocities, which were computed
by Lindblom and Ipser\cite{LI99} and presented in their Table 1. Our
above values have been obtained for $m=2$, but we can reproduce all
their values for $m=0$ to $m=3$.  Note, however, that their $l$ is
related to our $l_{\rm max}$ by $l = l_{\rm max} - 1$. Also note that
these values are completely independent of any stellar structure
parameters.

In Figs.~\ref{fig6} and \ref{fig7} we show both the continuous
spectrum and the eigenmodes as functions of the compactness $M/R$ for
different values of $l_{\rm max}$. The modes are for barotropic
stellar models, based on a polytropic equation of state with
$n=1$\cite{RSK02b}. Starting for $l_{\rm max} = 2$ (upper panel of
Fig.~\ref{fig6}), we find no mode, but only a band of frequencies
given by Eq.~(\ref{freqCowl}). We have also included the Newtonian
value for the $r$-mode as given by Eq.~(\ref{freqNewt}).  One can
clearly see that in the limit $M/R \rightarrow 0$ the continuous
spectrum shrinks to this value. If we include the coupling to the
polar equations with $l=3$, the situation changes drastically. As
expected we now obtain two continuous frequency bands and a mode which
lies below the Newtonian $r$-mode value, the further below the more
compact the star is.  Translating to positive frequencies, this means
that this mode always has a higher frequency than the Newtonian $r$
mode. Inspection of the eigenfunction reveals that the axial part is
well described by a $r^{l+1}$ power law, i.e.
\begin{eqnarray}
  u_3^{l=2} &\sim& r^3,
\end{eqnarray}
which is exactly what one expects for the $r$-mode. Inclusion of
higher $l$ only gives minor frequency corrections, which tells us that
this is indeed a physical inertial mode, the relativistic version of
an $m = 2$ $r$-mode. According to our nomenclature for inertial
modes, the $r$-mode is actually an $i_1^3$ inertial mode. In
Fig.~\ref{fig7}, we have also included the coupling up to $l_{\rm max}
= 4$ and 5. Here, each inclusion of one additional $l$ leads to
another patch in the continuous spectrum, and to the appearance of
some new inertial modes. Notice that the width of the uppermost and
lowermost patches is very broad, whereas the middle patches around
$\sigma/\Omega_K \approx -1$ remain very narrow, even for large $M/R$.
This is because the width of a patch is given by $\kappa\(\varpi_R -
\varpi_0\)$, which is therefore the smaller the closer $\kappa$ is
to zero. For any given $l_{\rm max}$, we find $l_{\rm max}-m$ new
inertial modes. We label them $i^l_n$ with $l = l_{\rm max}$ and $n =
1,\dots, l-m$. In the Newtonian limit, they appear very close to the
values found from $\det{\sf U} = 0$ for $l_{\rm max} - 1$. This is
what to be expected from the Newtonian theory of inertial modes,
since the eigenfrequency can be computed at lower order than the
eigenfunction.  This means that for the inertial modes $i_n^l$
appearing at $l_{\rm max} = l$, the associated eigenfrequencies can
already be inferred from the solutions of $\det{\sf U} = 0$ for
$l_{\rm max} - 1$.

From Fig.~\ref{fig6}, it is clear that the $r$-mode exists for the
whole compactness range, as it is always located outside the
continuous spectrum. For $l_{\rm max} = 4$ (upper panel of
Fig.~\ref{fig7}), it is still outside the continuous spectrum, but for
$l_{\rm max} = 5$ (lower panel of Fig.~\ref{fig7}), it lies inside the
continuous spectrum for $M/R > 0.14$. In this case, the eigenvalue
equations cannot be solved any more, as {\sf U} becomes singular. It
is only through the evolution of the time dependent equations that one
can see that the $r$-mode exists inside the continuous spectrum. This
is not only the case for the $r$-mode, but also for other inertial
modes. However, not all the modes are able to survive inside the
continuous spectrum. This can be seen for the inertial modes $i_1^4$
and $i_2^4$ in Fig.~\ref{fig7}.  For $l_{\rm max} = 4$, they both
cease to exist when they reach the continuous spectrum. For $l_{\rm
  max} = 5$ the continuous spectrum has shifted such that the $i_2^4$
mode remains always outside whereas the $i_1^4$ mode still cannot
penetrate the continuous spectrum. For larger $l_{\rm max}$, one can
find the $i_2^4$ mode inside the continuous spectrum, but not the
$i_1^4$ mode.

These results can be summarized as follows. As a result of the
relativistic frame dragging, there always exists a continuous
spectrum, which, however, depends very strongly on the number of
coupled equations. For $l_{\rm max} = 2$, the continuous spectrum is
confined to a single connected region, with increasing width for more
relativistic stellar models. If $l_{\rm max}$ is increased the
continuous spectrum splits into an increasing number of patches, which
are disconnected for weakly relativistic stars but tend to overlap for
more compact stellar models. In the limit $l_{\rm
  max}\rightarrow\infty$ the continuous spectrum should completely
cover the frequency range of the inertial modes. Nevertheless, there
are inertial modes, in particular the $r$-mode, which can exist inside
the continuous spectrum. For very relativistic stars, however, the
continuous spectrum is able to destroy some of the inertial modes.

Caution should be used in order to draw any conclusions about the
(non)existence of these modes for a rapidly rotating neutron star. The
above results showed that the continuous spectrum is very sensitive to
the number of equations that are coupled. So far, only first order
rotational corrections have been taken into account, i.e.~only the
coupling from $l$ to $l\pm1$. Inclusion of higher order corrections
leads to coupling to higher $l$, which can modify the continuous
spectrum in a significant way. It is therefore possible that higher
order corrections affect the continuous spectrum in such a way that
modes which do not exist in the first order analysis can now exist
because the responsible patch could have moved elsewhere.

Of course, another important step consists in including the metric
perturbations. Lockitch et al.\cite{LAF01} included some
(non-radiative) metric perturbations, but restricted their studies to
a post Newtonian treatment. Still, the continuous spectrum will remain
unaffected when metric variables are included, for it is only the
fluid equations which are responsible for its existence. However, by
including the metric perturbations, the modes might be affected in
such a way that they get pushed out of the continuous spectrum and we
might find some of the modes which we cannot find in the Cowling
approximation. That this might true can be seen from the purely axial
$l=m$ case, which we have discussed in the previous sections.  There,
the Cowling approximation leads to a purely continuous spectrum
without any mode solutions. In the low-frequency approximation, where
one retains a certain metric component, one is lead to Kojima's master
equation, which in certain cases admits $r$-mode solutions, whose
frequencies lie outside the continuous spectrum.  But the range of the
continuous spectrum is still the same as in the Cowling approximation
and even when all the metric perturbations were included, the
qualitative picture remained the same.

Nevertheless it is clear that the relativistic case is quite different
from the Newtonian one, and further studies are necessary to gain a
full understanding of the inertial modes of (rapidly) rotating
relativistic stars.

\section*{Acknowledgments}
We are grateful to N.~Andersson, E.~Berti, U.~Sperhake,
N.~Stergioulas, A.~Stavridis and M.Vavoulidis for many useful
comments and suggestions. This work has been supported by the EU
Programme 'Improving the Human Research Potential and the
Socio-Economic Knowledge Base' (Research Training Network Contract
HPRN-CT-2000-00137). J.R.~is supported by the Marie Curie
Fellowship No.~HPMF-CT-1999-00364.

\def\prl#1#2#3{{ Phys. Rev. Lett.\ }, {\bf #1}, #2 (#3)}
\def\prd#1#2#3{{ Phys. Rev. D}, {\bf #1}, #2 (#3)}
\def\prep#1#2#3{{ Phys. Reports}, {\bf #1}, #2 (#3)}
\def\jmp#1#2#3{{ J. Math. Phys.}, {\bf #1}, #2 (#3)}
\def\cqg#1#2#3{{ Class. Quantum Grav.}, {\bf #1}, #2 (#3)}
\def\apj#1#2#3{{ Astrophys. J.}, {\bf #1}, #2 (#3)}
\def\apjl#1#2#3{{ Astrophys. J. Lett.}, {\bf #1}, #2 (#3)}
\def\apjs#1#2#3{{ Astrophys. J. Suppl.}, {\bf #1}, #2 (#3)}
\def\acta#1#2#3{{ Acta Astronomica}, {\bf #1}, #2 (#3)}
\def\aa#1#2#3{{ Astron. Astrophys.}, {\bf #1}, #2 (#3)}
\def\apss#1#2#3{{Astrop. Sp. Sci.}, {\bf #1}, #2 (#3)}
\def\mnras#1#2#3{{ Mon. Not. R. Astr. Soc.}, {\bf #1}, #2 (#3)}
\def\prsla#1#2#3{{ Proc. R. Soc. London, Ser. A}, {\bf #1}, #2 (#3)}
\def\ijmpd#1#2#3{{ I.J.M.P.} D {\bf #1}, #2 (#3)}
\def\aa#1#2#3{{ A\&A}  {\bf #1}, #2 (#3)}
\def\ptp#1#2#3{{ Prog. Theor. Phys.}  {\bf #1}, #2 (#3)}
\def\nature#1#2#3{{ Nature}  {\bf #1}, #2 (#3)}
\def\cmp#1#2#3{{ Commun. Math. Phys.}  {\bf #1}, #2 (#3)}
\def\lrr#1#2#3{{ Living Rev. Rel.}  {\bf #1}, #2 (#3)}

\end{document}